\documentclass{aa}  
\usepackage{graphicx}
\usepackage{txfonts}
\usepackage{hyperref}

\begin{document}

   \title{Magnetar-like flares behind the high-energy emission in LS~5039}

   \author{V. Bosch-Ramon
          \inst{1}
          \and
          M.V. Barkov\inst{2}
          }

   \institute{Departament de Física Quàntica i Astrofísica, Institut de Ciències del Cosmos, Universitat de Barcelona, IEEC-UB, Martí i Franquès 1, 08028 Barcelona, Spain\\
              \email{vbosch@fqa.ub.edu}
         \and
             Institute of Astronomy, Russian Academy of Sciences, Moscow, 119017 Russia\\
             \email{barkov@inasan.ru}             
             }

   \date{Received November 24, 2024; accepted ...}

  \abstract
   {LS~5039 is a system hosting a high-mass star and a compact object of unclear nature. There are hints that the system may host a strongly magnetized neutron star, a scenario that requires a mechanism to power its persistent and strong nonthermal emission.}
   {We investigate a mechanism in which the nonsteady interaction structure of the stellar and the compact object winds can regularly excite neutron star magnetospheric activity, which can release extra energy and fuel the source nonthermal emission.}
   {The neutron star wind shocked by the stellar wind can recurrently touch the neutron star magnetosphere, triggering magnetic instabilities whose growth can release extra energy into the neutron star wind in a cyclic manner. To illustrate and study the impact of these cycles on the two-wind interaction structure on different scales, we performed relativistic hydrodynamics simulations in two and three dimensions with periods of an enhanced power in the neutron star wind along the orbit. We also used analytical tools to characterize processes near the neutron star relevant for the nonthermal emission.}
   {As the neutron star wind termination shock touches the magnetosphere energy dissipation occurs, but the whole shocked two-wind structure is eventually driven away, stopping the extra energy injection. However, due to the corresponding drop in the neutron star wind ram pressure, the termination shock propagates back toward the magnetosphere, resuming the process. These cycles of activity excite strong waves in the shocked flows, intensifying their mixing and the disruption of their spiral-like structure produced by orbital motion. Further downstream, the shocked winds can become a quasi-stable, relatively smooth flow.}
   {The recurrent interaction between the neutron star magnetosphere and a shocked wind can fuel a relativistic outflow powerful enough to explain the nonthermal emission of LS~5039. A magnetospheric multipolar magnetic field much stronger than the dipolar one may provide the required energetics, and help to explain the lack of evidence of a recent supernova remnant.}

   \keywords{Stars: magnetars -- Hydrodynamics -- X-rays: binaries -- stars: winds, outflows -- radiation mechanisms: nonthermal}

   \maketitle

\section{Introduction}\label{intro}

LS~5039 is a binary system hosting an O6.5~V star and a compact object of unclear origin \citep{cas05}, and a powerful gamma-ray source with a nonthermal radiation luminosity, $L_{NT}\sim 10^{36}$~erg~s$^{-1}$, peaking around $\sim 10-100$~MeV \citep[e.g.,][]{col14}. As was shown in \cite{cas05}, a black hole in the system is likely only if the latter is pseudo-synchronized, although hints of $P\approx 9$~s period pulsations found in X-rays would favor a pulsar if confirmed \citep{yon20}. Despite evidence of X-ray pulsations being under debate \citep{kar23,mak23}, they would strongly indicate the presence of a slowly rotating, magnetar-like neutron star (NS), make the standard scenario of a pulsar in the ejector regime less likely due to a low wind power, and favor a magnetospheric origin for the energy fueling the nonthermal emission \citep{yon20}. It is worth noting that LS~I~+61~303, a similar high-mass binary and gamma-ray emitter, has been found to host a relatively slowly rotating NS that may also present magnetar-like behavior \citep[e.g.,][and references therein]{wen22}. Given all this, a model for energizing the nonthermal emitter with the energy stored in the NS magnetosphere may be required in these sources. 

\cite{yon20} explored analytically the possibility of feeding the nonthermal emission in LS~5039 with energy from the magnetosphere, released due to its interaction with the stellar wind, but the specifics of this interaction were not provided. Moreover, the presence of a magnetar in the system should be reconciled with the absence of evidence for a very young supernova remnant (SNR) near LS~5039 \citep[][]{mol12}. Magnetar flares have been discussed in the literature \citep[see, e.g.,][and references therein]{2020ApJ...896..142B,2022ApJ...927....2K} as the sources of fast radio bursts \citep[FRBs;][]{2007Sci...318..777L}, particularly in the context of flares of a magnetar in a high-mass binary \citep{boz08} when there is periodicity \citep[see, e.g.,][]{2020ApJ...893L..39L, 2022MNRAS.515.4217B}, so LS~5039 hosting a magnetar would fit in this framework. Thus, in this work we propose a mechanism for LS~5039 based on recurrent periods of magnetar-like flaring activity. These periods start with a weak NS (or pulsar) wind that cannot balance the ram pressure of the wind from the companion star, and the shocked pulsar wind can reach or even enter the light cylinder of the NS, directly touching its magnetosphere (as is proposed in \citealt{yon20}). This can trigger magnetic perturbations in the magnetosphere that can quickly grow, and ultimately lead to magnetic reconnection and the release of large amounts of energy into the otherwise weak pulsar wind beyond the light cylinder \citep{bar22,2023MNRAS.524.6024S}. The injection of magnetospheric energy lasts as long as the pulsar wind termination shock touches the magnetosphere, and during this period the shocked pulsar wind mediates the energy transfer between the latter (i.e., the energy source) and the region resulting from the star-pulsar wind interaction (i.e., where -or downstream of which- the nonthermal emitter is located). Eventually, though, the enhanced pulsar wind drives the whole shocked flow structure beyond the light cylinder, stopping the injection of energy. This severely reduces the pulsar wind ram pressure and the shocked pulsar wind quickly expands inward until it again touches the magnetosphere, moment at which the cycle can repeat. This process leads to recurrent periods of power-enhanced pulsar wind that feed the overall nonthermal emission in the system. To show how the shocked structure evolves, we carried out relativistic hydrodynamics (RHD) simulations that mimic the effect of the magnetic active periods on the shocked winds, including the role of orbital motion. We also explored, using analytical tools, the processes near the NS relevant for the nonthermal emission. 

Regular pulsar winds (i.e., not enhanced by magnetar activity) can already be significantly magnetized and nonspherically symmetric \citep[e.g.,][]{2019MNRAS.490.3601B}. Moreover, the part of the wind enhanced by magnetar activity can also cover a relatively narrow solid angle of a few steradians, which increases the wind anisotropy \citep[][]{2023MNRAS.524.6024S}. In addition, three-dimensional (3D) relativistic magnetohydrodynamics (RMHD) simulations \citep[][]{2014MNRAS.438..278P,2024PASA...41...48B} showed that the shocked pulsar wind structure is very sensitive to changes in the environment. Thus, the magnetic field is indeed a major factor in the problem, and should eventually be included in simulations of the discussed scenario. However, the 3D RMHD solution turned out to be on average similar to that for a spherical RHD wind, which also shows strong instabilities \citep[see sect.~3 in][]{2024PASA...41...48B}. Furthermore, the simulations presented here took about 100~000~CPU~hours; since an RMHD simulation requires at least a 10 times higher spatial resolution, the computational time should be $\sim10^9$~CPU~hours, which is extremely hard to achieve. 
For all this, at this stage RHD simulations were employed as they provide qualitatively relevant results, while keeping the computing costs at a reasonable level and allowing for an exploratory study.

The article is structured as follows: In Sect.~\ref{sec:setup} the simulations are described, and their results are shown in Sect.~\ref{sec:res}. Then, in Sect.~\ref{disc}, a discussion is carried out about the regime of interaction between the NS and the surrounding medium, the origin of the nonthermal energy and the age of the system, and the nonthermal emitter.

\section{Simulation setup}
\label{sec:setup}

The RHD simulations were performed in 2D and 3D geometry in Cartesian (2D and 3D) and cylindrical (axisymmetric 2D) coordinates using the PLUTO code\footnote{Link http://plutocode.ph.unito.it/index.html} \citep{2007ApJS..170..228M}. 
Spatial linear interpolation, a third-order Runge-Kutta approximation in time, and an HLLC Riemann solver were adopted \citep{2005JCoPh.203..344L}. PLUTO is a modular Godunov-type code entirely written in C and intended mainly for astrophysical applications and high-Mach-number flows in multiple spatial dimensions. The simulations were performed on the CFCA XC50 cluster of the National Astronomical Observatory of Japan (NAOJ) and RIKEN HOKUSAI Bigwaterfall. The flow was approximated as an ideal, relativistic adiabatic gas, with one particle species and a polytropic index of 4/3. The size of the Cartesian domain is $x$ and $y\in[-32, 32]$ in 2D and 3D, and $z\in [-20, 20]$ in 3D. The resolution is nonuniform in the entire computational domain, with a total number of cells: $N_{X} = N_{Y}= 780$ (2D and 3D), and $N_{Z}  = 650$ (3D; see details in Table~\ref{tab:grid}). The size of the axisymmetric domain is $r\in[0, 0.7]$, and $z\in [0.7, 2.0]$, with a total number of cells: $N_{r} =  260$, and $N_{Z}  = 360$. The higher-resolution central part features $N_{r} =  130$, and $N_{Z}  = 260$, with size $r\in[0, 0.25]$, and $z\in [0.7, 1.2]$.

In this paper, the unit of length $a = 2.2\times 10^{12}$~cm corresponds to the semimajor axis for an orbital period of $T_{orb}\approx 3.9$~days and a binary total mass of $\approx 30$~M$_\odot$ \citep{cas05}. This work was intended as a first exploration of the phenomenon under study. Thus, the results of the Cartesian 2D simulations were meant to be illustrative of the large-scale evolution \citep{2012A&A...544A..59B}, but the difficulty in capturing in detail the magnetospheric scale ($\sim 10^{-2}$ times smaller than $a$; see Sect.~\ref{disc}) precluded in that case a detailed characterization of the shocked pulsar wind evolution right between the star and the NS. The latter limitation also affects the Cartesian 3D simulations. Nevertheless, the tendencies in the flow evolution on the smallest captured scales hinted at the role of the shocked pulsar wind when enhanced by strong magnetospheric activity. In the Cartesian simulations, $a$ corresponds to $\approx 64$~cells. On the other hand, the axisymmetric 2D simulations allowed us to study the shocked two-wind region closest to the magnetosphere, with $a$ corresponding now to $\approx 520$~cells, but without including orbital motion, the latter playing a minor role on those scales.
Finally, the equation of state adopted, albeit simple, is already enough to derive meaningful conclusions from the simulations \citep[see the explorations carried out in][]{2012A&A...544A..59B,2015A&A...577A..89B,2024PASA...41...48B}.

\begin{table*}
\caption{Parameters of the Cartesian grid (“left” extended grid, central grid, “right” extended grid; see text for details -Sect.~\ref{sec:setup}-).}
\begin{tabular}{lccccccc}
\hline
\hline
   Coordinates        &  Left~($\times a$)    &  $N_l$   &   Left-center~($\times a$)  &   $N_c$ &   Right-center~($\times a$) & $N_r$ & Right~($\times a$)  \\
\hline
  3D case         &      &     &    &    &   &  &   \\
\hline
&&&&&\\[-5pt]
 $ x  $     & \quad $-32$ & \quad $260$ & \quad $-2$ &\quad 260 & \quad $2$ &\quad 260 & \qquad $32$ \\
$ y  $     & \quad $-32$ & \quad $260$ & \quad $-2$ &\quad 260 & \quad $2$ &\quad 260 & \qquad $32$ \\
 $ z  $     & \quad $-20$ & \quad $260$ & \quad $-1$ &\quad 130 & \quad $1$ &\quad 260 & \qquad $20$ \\
\hline
  2D cases       &      &     &    &    &   &  &   \\
\hline
 $ x  $     & \quad $-32$ & \quad $260$ & \quad $-2$ &\quad 260 & \quad $2$ &\quad 260 & \qquad $32$ \\
$ y  $     & \quad $-32$ & \quad $260$ & \quad $-2$ &\quad 260 & \quad $2$ &\quad 260 & \qquad $32$ \\
\hline
&&&&&\\[-5pt]
\end{tabular}
\label{tab:grid}
\end{table*}

\subsection{Setup of the winds}
\label{sec:setwind}

The wind of the massive star is assumed to be cold and move radially with constant velocity $v_w=2400$~km/s and Mach number $M_w = 7$. The pulsar wind is also cold and moves radially with Lorentz factor $\Gamma_{pw}=1.9$ and Mach number $M_{pw}=17$, and the star-to-pulsar wind momentum rate ratio is $\eta=0.05$ in the weak wind phase in the Cartesian simulations and $\eta=4\times 10^{-4}$ in the axisymmetric simulations, where 
\begin{equation}
    \eta = \frac{\dot E_{sd}}{c\dot{M}v_w},
    \label{eq:eta}
\end{equation}
with $\dot E_{sd}$ being the spin-down power of the pulsar and $\dot{M}$ the massive star mass-loss rate. The winds were modeled similarly to those in \cite{2015A&A...577A..89B}.
During flaring periods, in the Cartesian case $\eta$ grows such that its orbit-averaged value reaches 0.16, but in one case it reaches 0.39 (in the axisymmetric case only one flare is presented; see the next section).

\subsection{Setup of the magnetar flaring periods}
\label{sec:setflares}

\begin{table*}
\caption{Parameters of the Cartesian models.
}
\begin{tabular}{lcccccccc}
\hline
\hline
  Name     & Dimen.  &  $N_{fl}$    & $A_{fl}$ & (<$\dot{E}_{fl}$>+<$\dot{E}_{sd}$>)/<$\dot{E}_{sd}$>  & $\Delta$ & $\eta$ & $\max(\eta_{fl})$ & <$\eta$> \\
\hline
&&&&&&&&\\[-5pt]
 3Dn30e10  & \quad 3D   & \quad $30$ & \quad $10$ & \quad $3.3^*$ & $0.05$ & $0.05$ & $3.6$ &  $0.16^*$ \\
\hline
&&&&&&&&\\[-5pt]
 2Dn10e10     & \quad 2D & \quad $10$  & \quad $10$ & \quad $3.3$ & $0.05$ & $0.05$ & $3.6$ & $0.16$ \\
 2Dn30e10     & \quad 2D & \quad $30$  & \quad $10$ & \quad $3.3$  & $0.05$ & $0.05$ & $3.6$ & $0.16$\\
 2Dn90e10     & \quad 2D & \quad $90$  & \quad $10$ & \quad $3.3$  & $0.05$ & $0.05$ & $3.6$ & $0.16$\\
 2Dn30e30     & \quad 2D & \quad $30$  & \quad $30$ & \quad $7.8$  & $0.05$ & $0.05$ & $10.7$ & $0.39$\\
\hline
 nfeta0.05    & \quad 2D & \quad $0$ & \quad $0$ & \quad $1$  & N/A & $0.05$ & $0.05$ & $0.05$\\
 nfeta0.5     & \quad 2D & \quad $0$ & \quad $0$ & \quad $1$  & N/A & $0.5$  & $0.5$  & $0.5$ \\
\hline
&&&&&&&&\\[-5pt]
\end{tabular}
\tablefoot{Parameter symbols: $N_{fl}$ (number of flares per orbit), $A_{fl}$ (flare amplitude factor); (<$\dot{E}_{fl}$>+<$\dot{E}_{sd}$>)/<$\dot{E}_{sd}$> (orbit-averaged energy enhancement); $\Delta$ (flare relative width), $\eta$ (pulsar-star wind momentum rate ratio); $\max(\eta_{fl})$ (maximum $\eta$ within a flare); <$\eta$> (orbit-averaged $\eta$; $^*$ in 3D, average over the covered orbit fraction).}
\label{tab:models}
\end{table*}

\begin{figure}
   \centering 
   \includegraphics[width=\hsize]{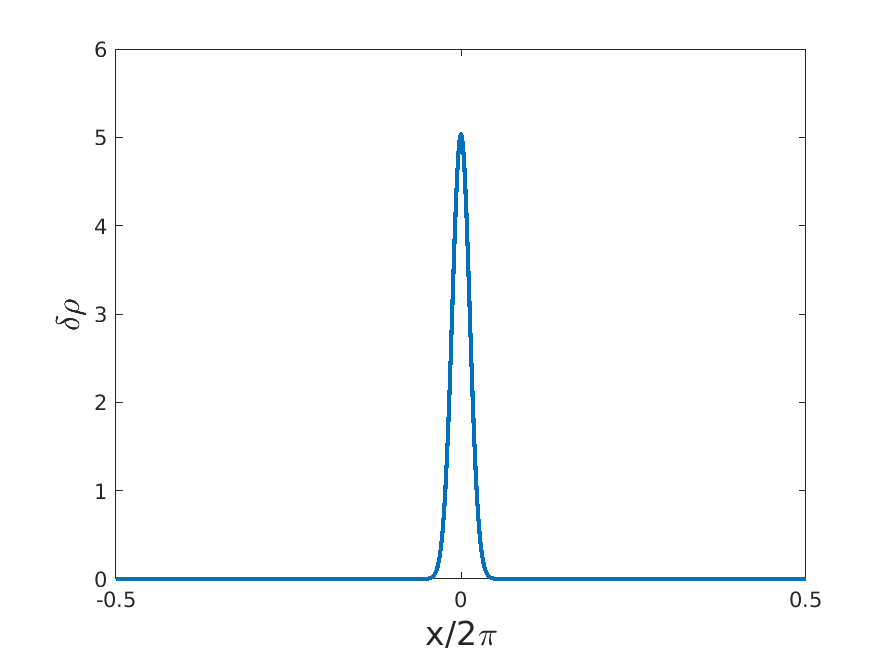}
   \caption{Example of the time profile of a flaring period; $\delta\rho$ is the density on top of the regular pulsar wind density during flaring periods, $x= 2\pi t N_{fl}/T_{orb}$ the dimensionless time (see text for details -Sect.~\ref{sec:setwind}-), and $\Delta =0.1$ the width of the pulse.}
   \label{fig:imp}
\end{figure}

The flaring periods were approximated via a function representing recurrent additional injections of power,
$\dot E_{pw}\propto \dot E_{sd} (1+\delta\rho) \propto \rho_{pw} (1+\delta\rho)$,
which has the following time profile:
\begin{equation}
    \delta\rho = A_{fl}\frac{5}{\pi}\frac{e^{-\bar y}}{\Delta^{1/2}}\,,\quad \mbox{ with }\quad \bar y\equiv 15.917\frac{1-\cos{\bar x}}{\Delta}\,,
    \label{eq:drho}
\end{equation}
where $\bar x = 2\pi t N_{fl}/T_{orb}$ is a dimensionless time variable normalized to the orbital period, $T_{orb}$, $\rho_{pw}$ is the weak pulsar wind density, $N_{fl}$ the number of flaring periods per orbit, $\Delta$ the relative width of the flaring period, and $A_{fl}$ its amplitude. The last two terms quantify how much energy per orbit is released by the flaring periods compared to the one from the weak pulsar wind alone, $T_{orb}$<$\dot E_{fl}$> and $T_{orb}$<$\dot E_{sd}$>, respectively\footnote{We chose the normalization parameters to satisfy the following equation $\int_{-\pi}^{\pi}\delta\rho(\bar x) d\bar x = 1$ with an accuracy better than 1\%.}. The time profile of the flaring periods was chosen to be sharp, while avoiding numerical artifacts in the simulations. A profile with $\Delta = 0.1$ is presented in Fig.~\ref{fig:imp} for better visualization, but in the simulations a narrower profile with $\Delta = 0.05$ was used. In the Cartesian simulations, the energy injection averaged over the orbit including both the flaring and weak wind periods releases 3.3 ($\equiv<\eta>=0.16$) and 7.8 ($\equiv<\eta>=0.39$) times more energy than the weak pulsar wind alone ($\eta=0.05$). The maximum enhancement of injected power within the flaring periods is higher: 72 ($\max(\eta_{fl})=3.6$) and 214 ($\max(\eta_{fl})=10.7$) times larger than in the weak wind phase. The parameter values used are presented in Table~\ref{tab:models}.
In the axisymmetric simulations, we were exploring a different spatial and temporal scale, as was allowed by the more focused setup, and the flare was made significantly shorter to mimic the response of the shocked structure on the scales of the magnetospheric radius. Thus we took $\Delta = 0.003$,
and the adopted maximum enhancement of injected power during the flaring period was higher, by a factor of 870, which corresponds to $\max(\eta_{fl})=0.35$ from $\eta = 4\times10^{-4}$.

\section{Results}
\label{sec:res}

   \begin{figure*}
   \centering
   \includegraphics[width=88mm]{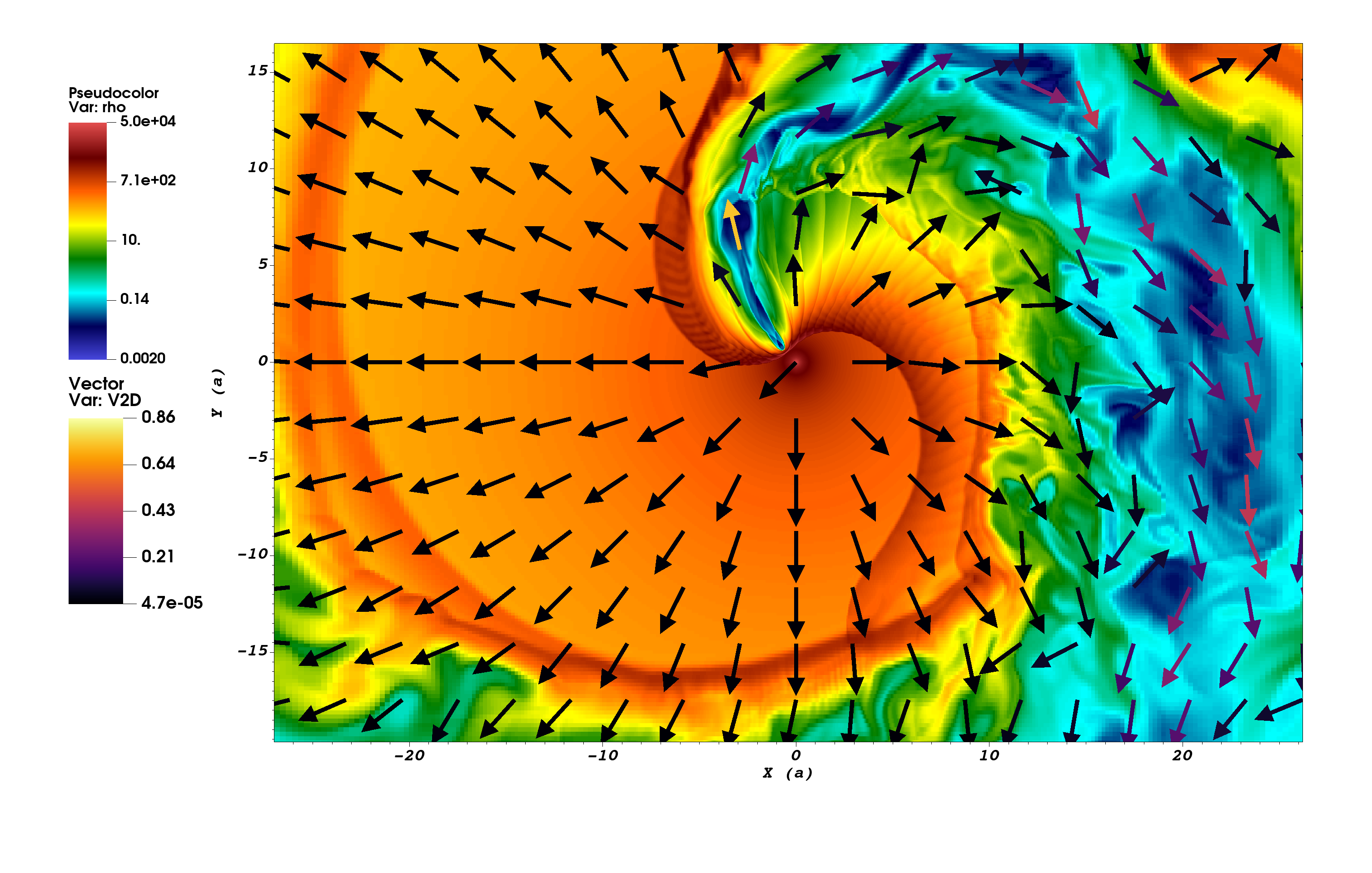}
   \includegraphics[width=88mm]{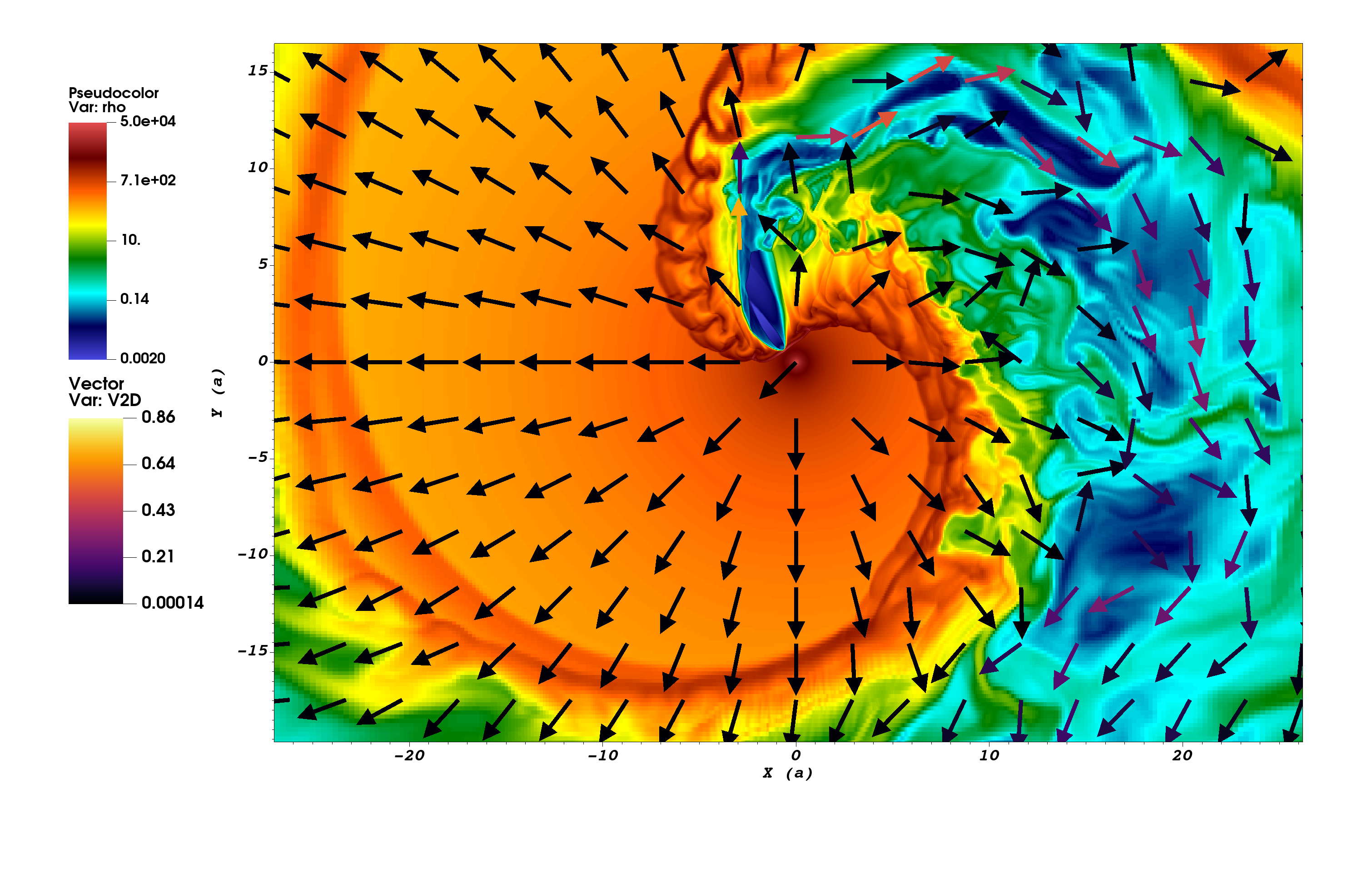}
   \includegraphics[width=88mm]{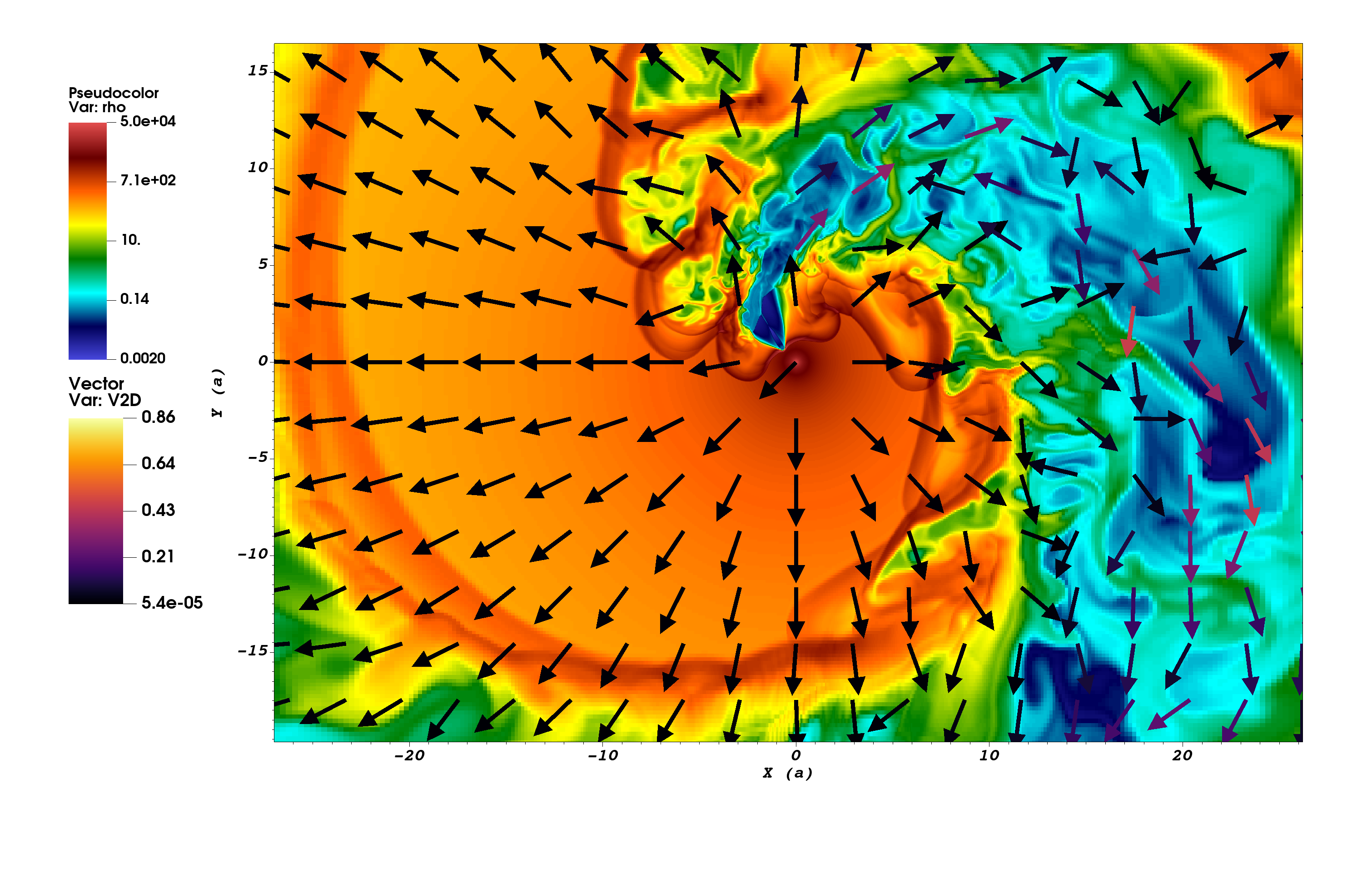}
   \includegraphics[width=88mm]{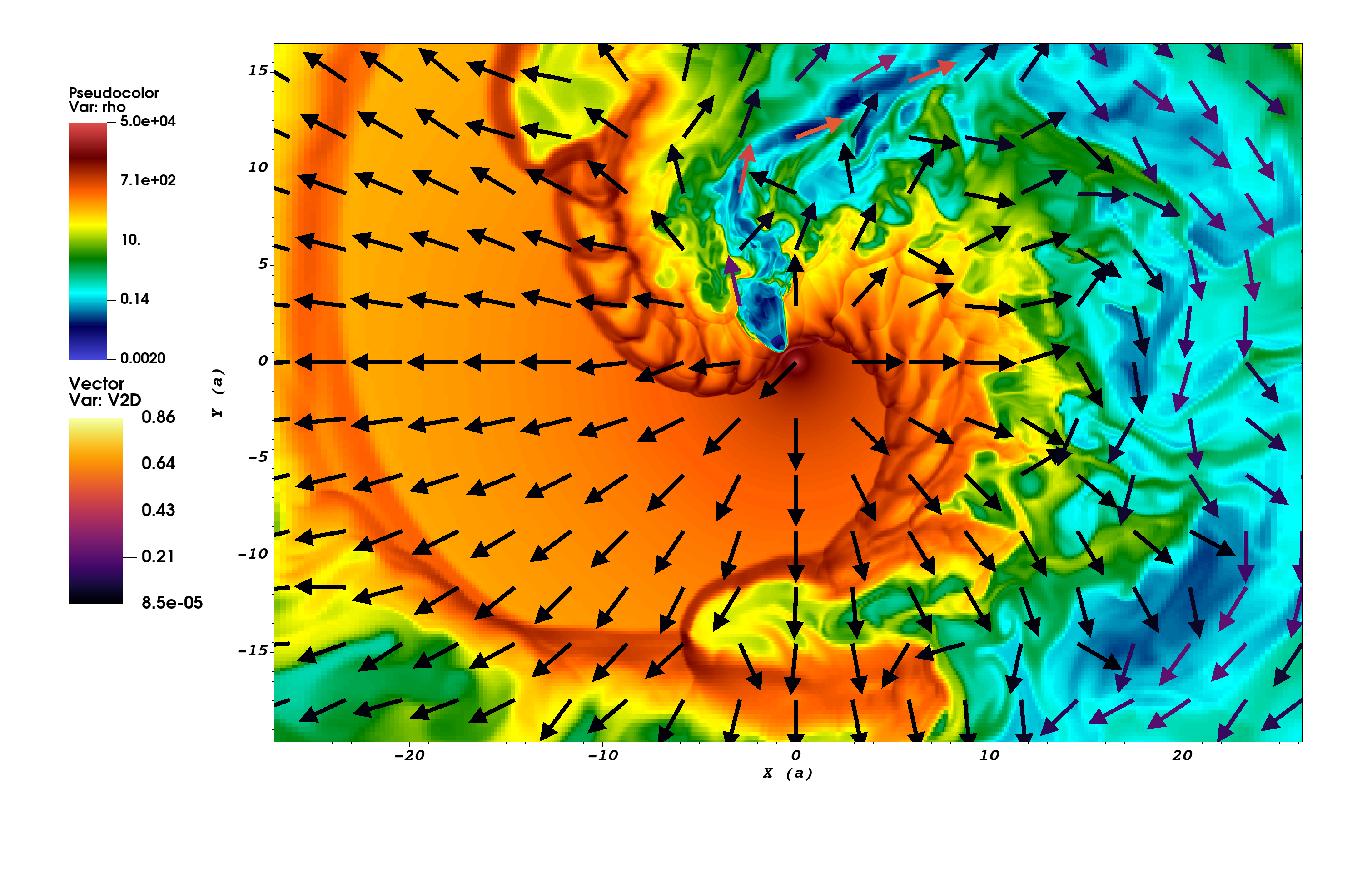}
   \includegraphics[width=88mm]{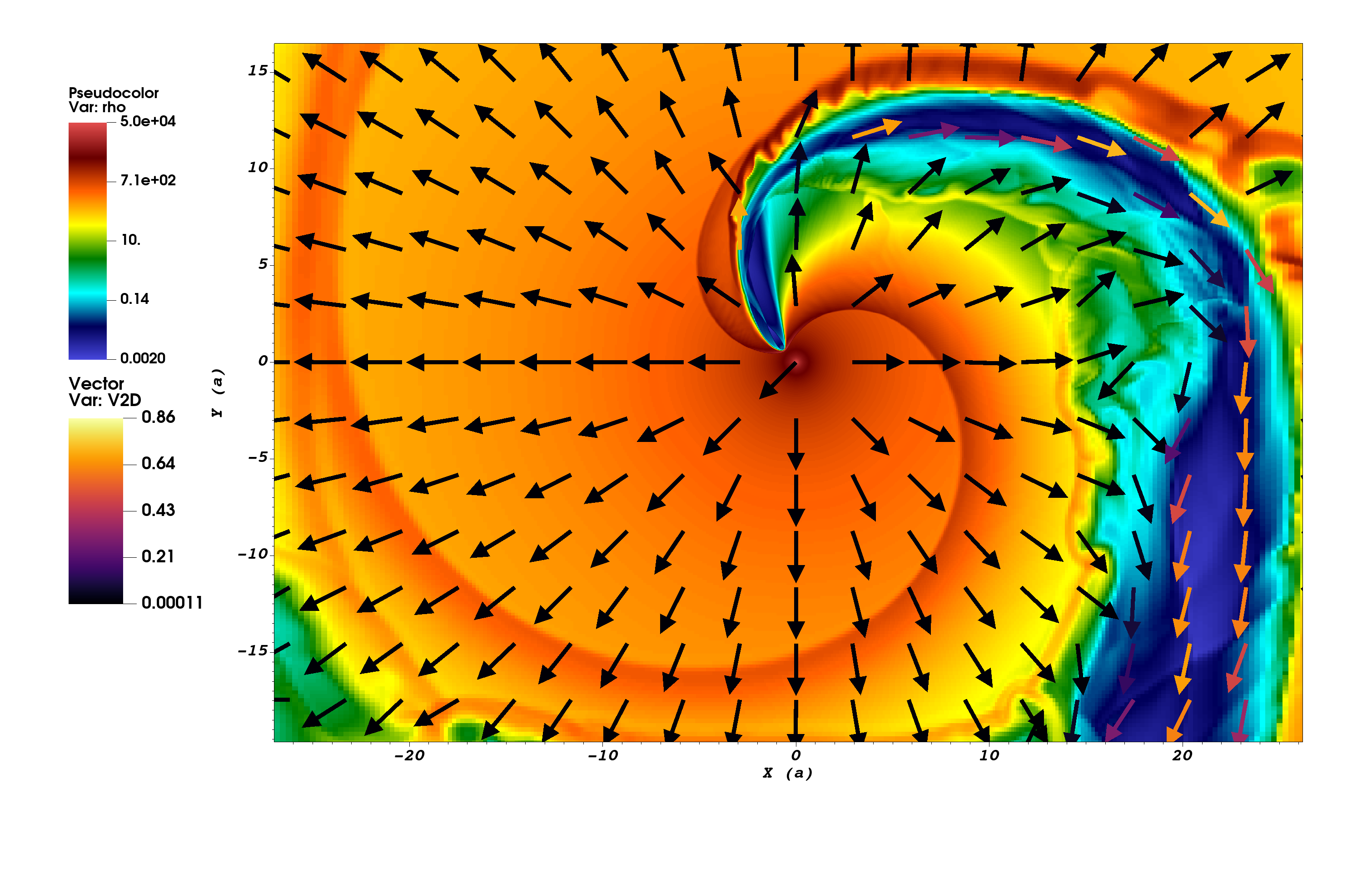}
   \includegraphics[width=88mm]{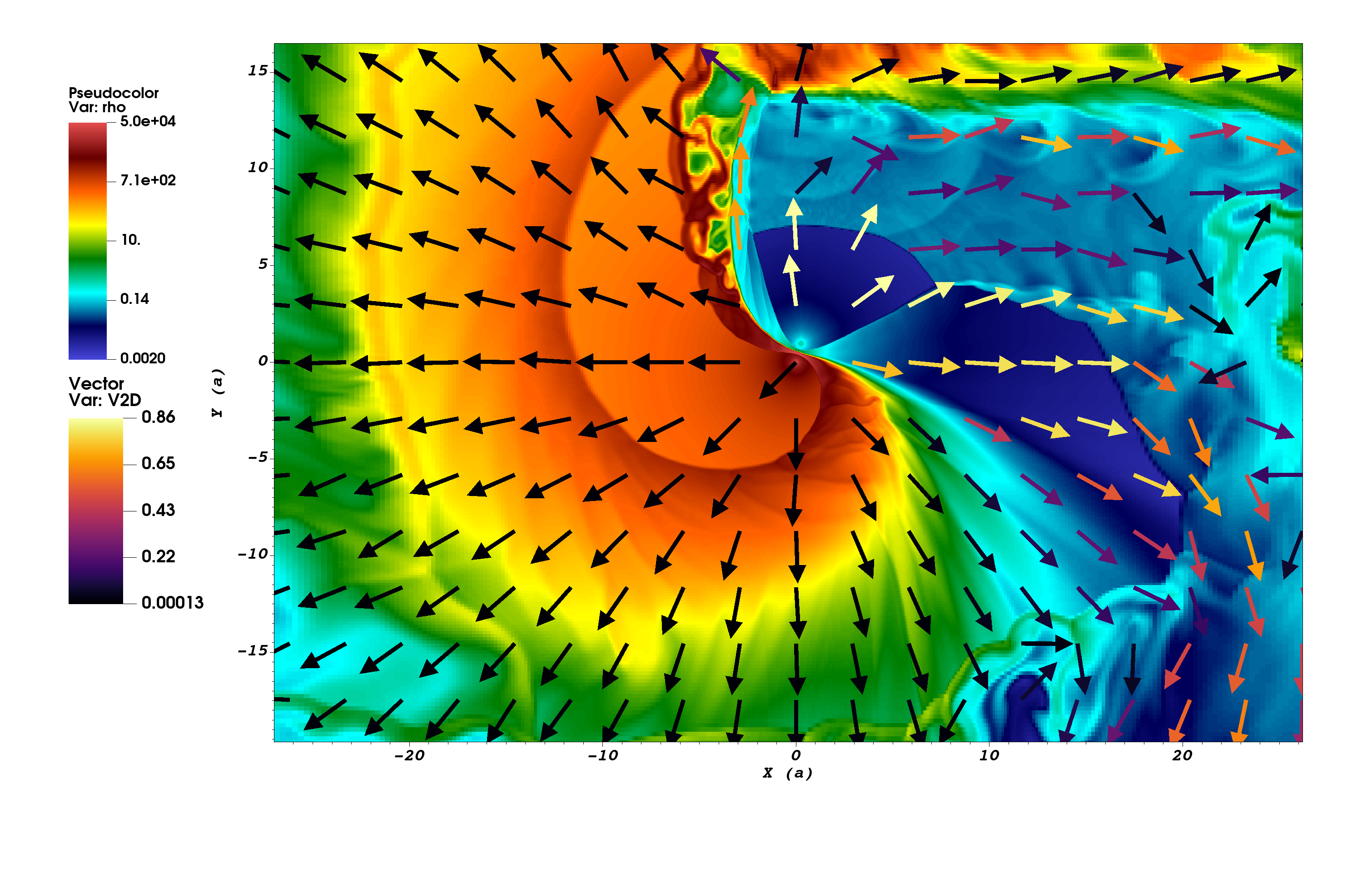}
   \vspace{-0.75cm}
   \caption{Colored density maps on the orbital plane, with colored arrows representing the modulus of the 3-velocity vector for the 2D models. The flaring period rates are $N_f=90$ (top left, 2Dn90e10), 30 (top right, 2Dn30e10), and 10 (middle left, 2Dn10e10) flaring periods per orbit, keeping the total energy budget of the system constant. In the model 2Dn30e30, we increased the power of the flaring period by a factor of 3, while keeping $N_f= 30$ (middle right panel). The non-flaring cases are presented at the  bottom left (nfeta0.05) and  right (nfeta0.5).}
    \label{fig:rhov2D}
    \end{figure*}

   \begin{figure}
   \centering
   \includegraphics[width=88mm]{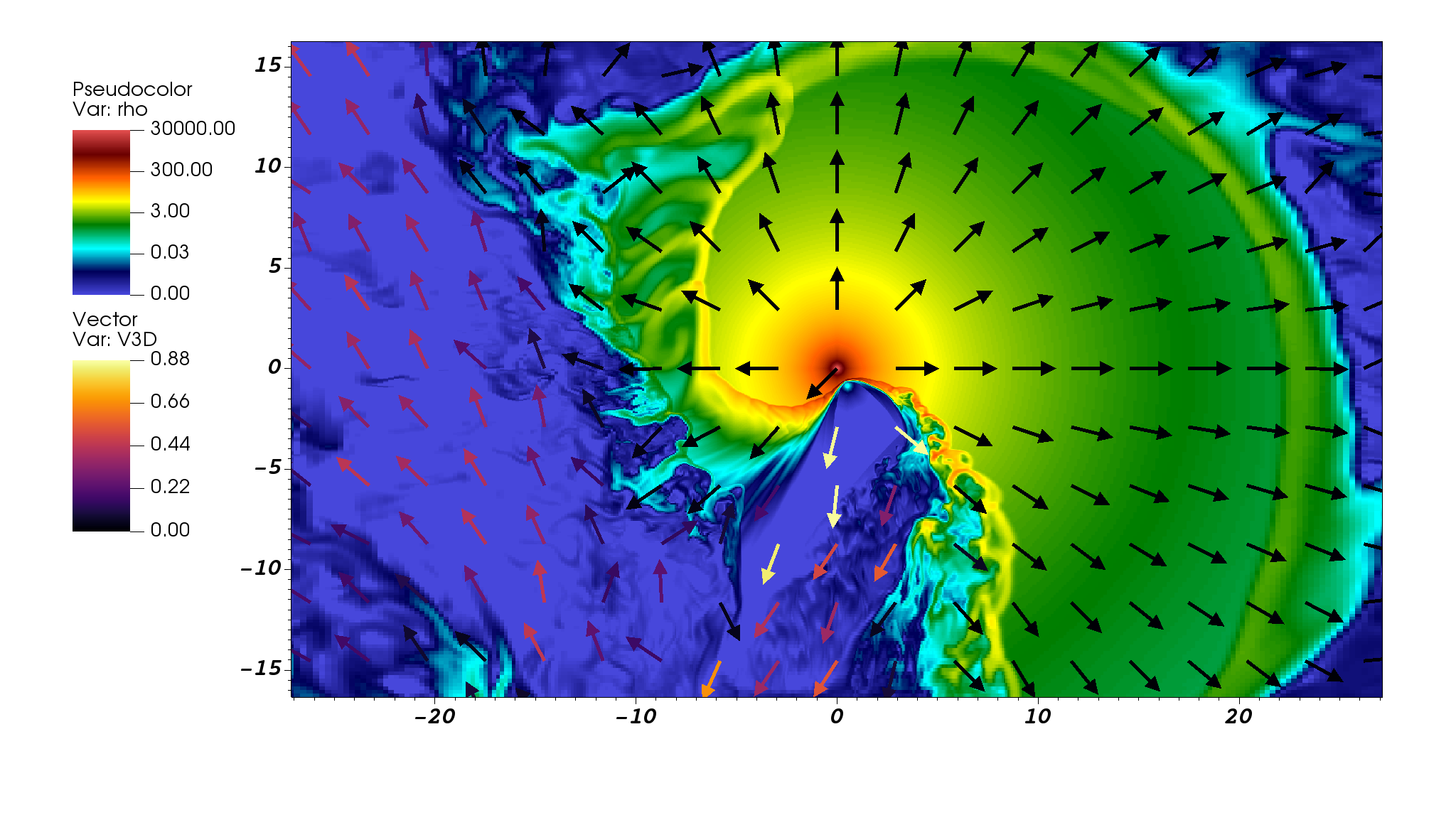}
   \includegraphics[width=88mm]{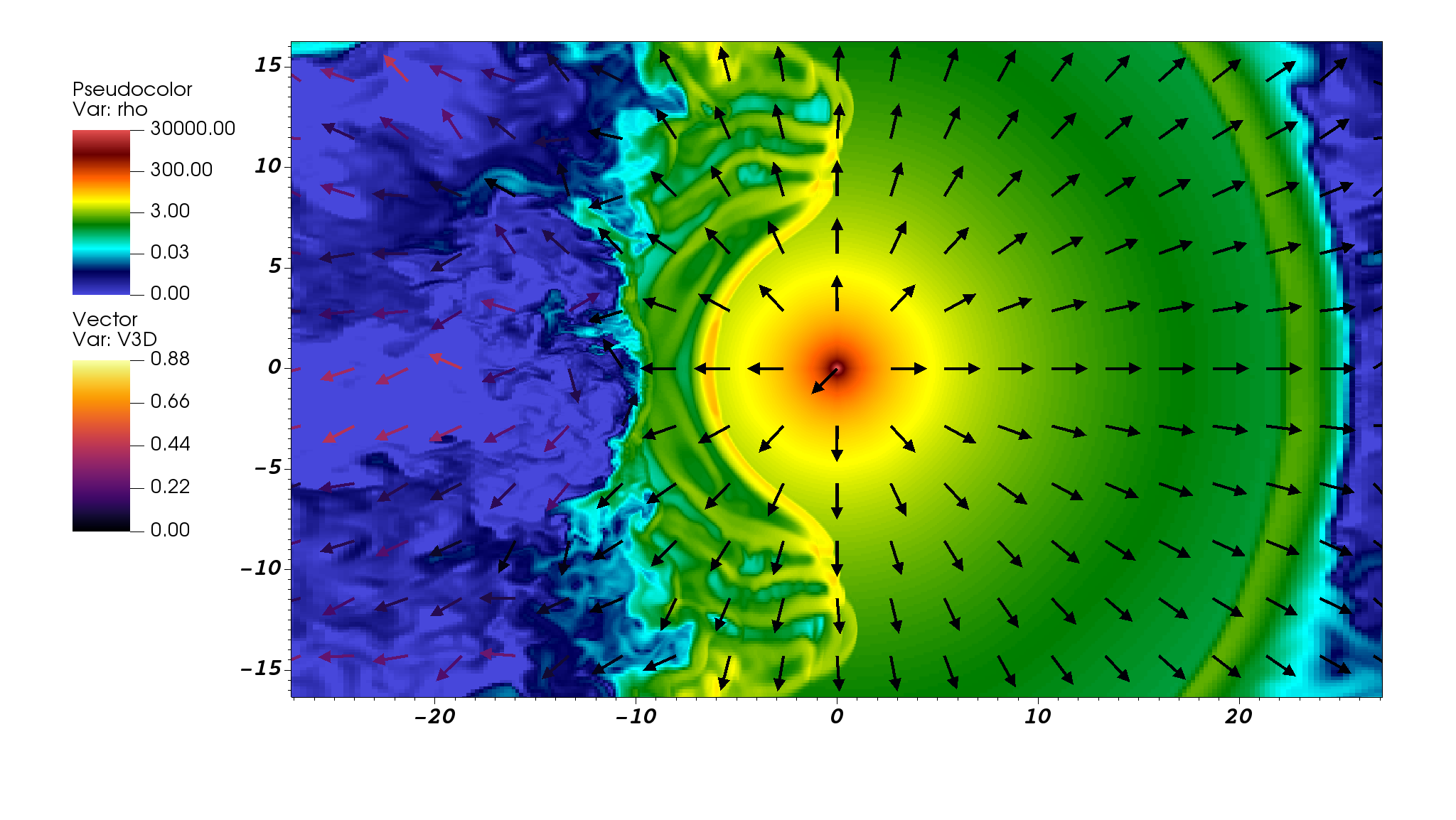}
   \includegraphics[width=88mm]{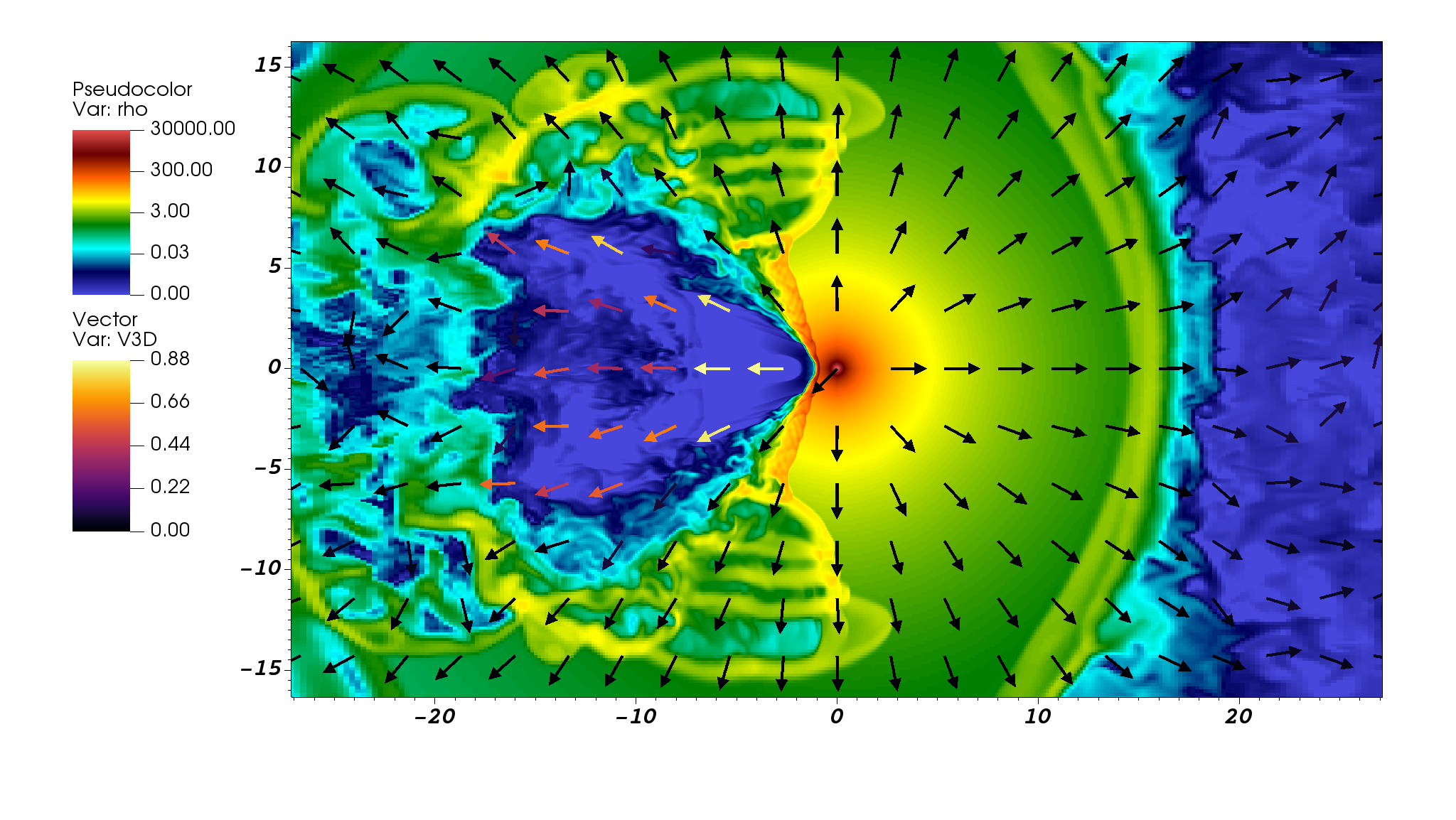}
   \vspace{-0.5cm}
   \caption{Colored density maps with colored arrows representing the modulus of the 3-velocity vector for the 3D model (3Dn30e10) for various cuts: $XY$ (top; orbital plane), $XZ$ (middle), and $ZY$ (bottom).}
    \label{fig:rhov3D}
    \end{figure}    

   \begin{figure*}
   \centering
   \includegraphics[width=88mm]{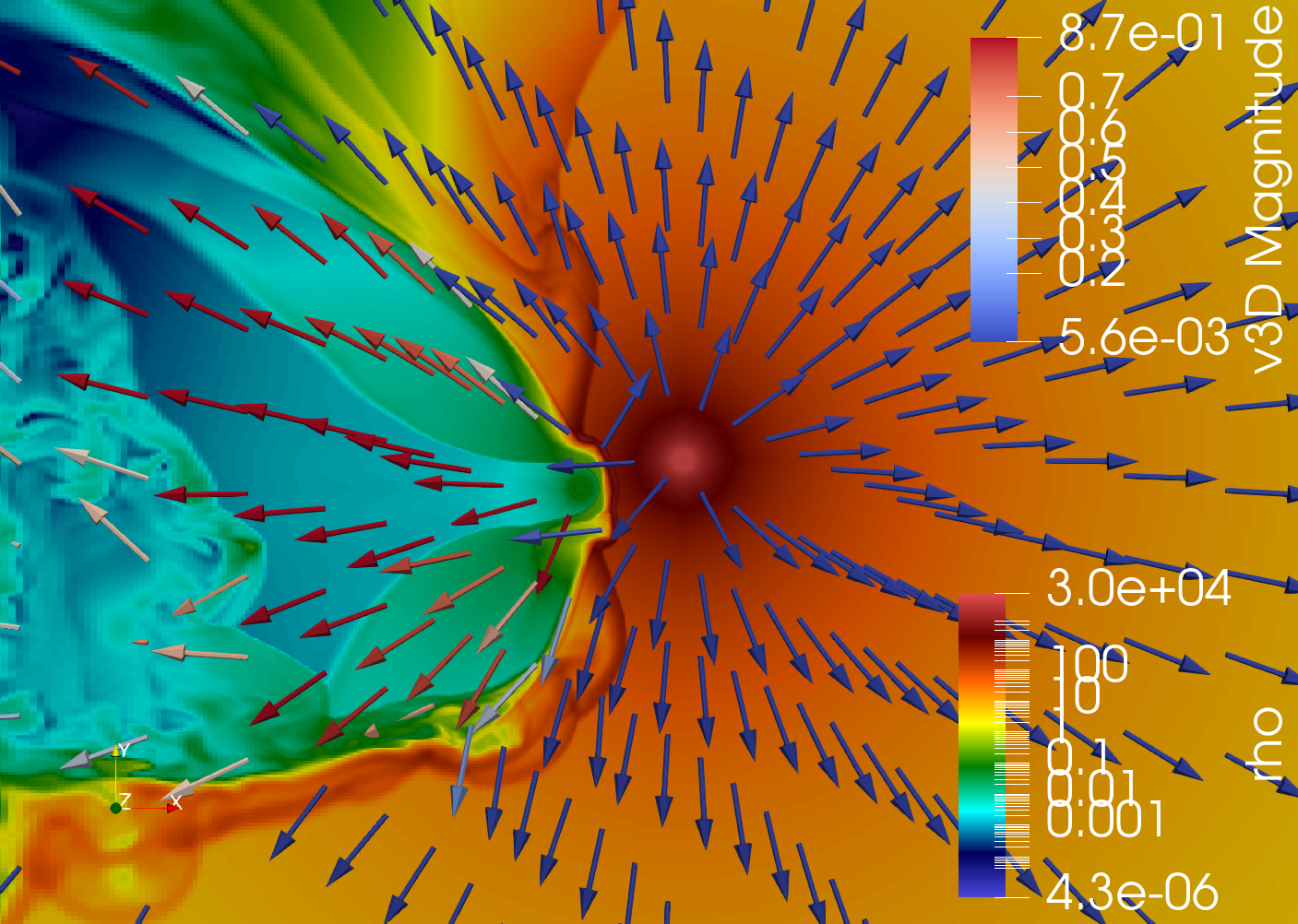}
   \includegraphics[width=88mm]{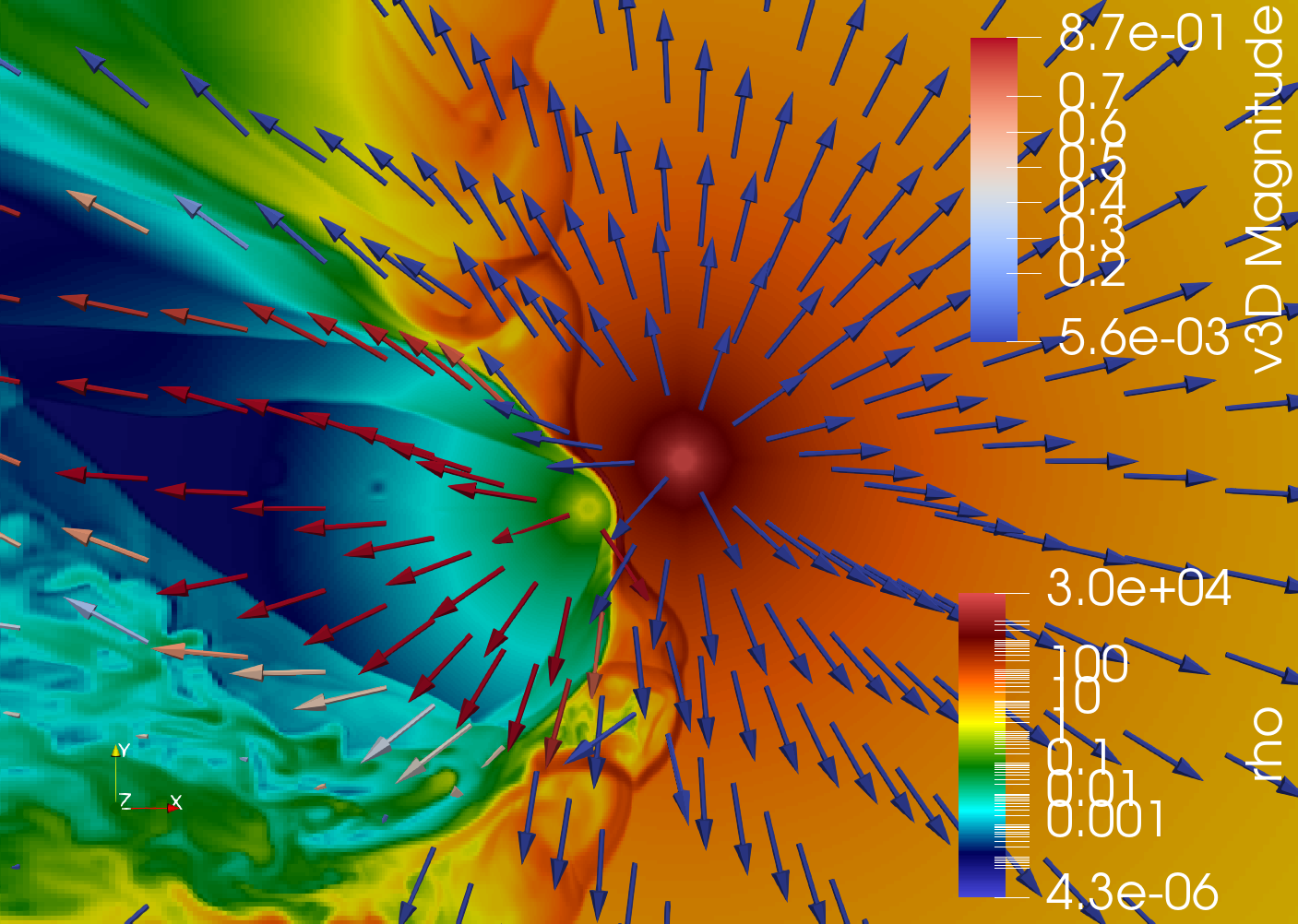}
   \includegraphics[width=88mm]{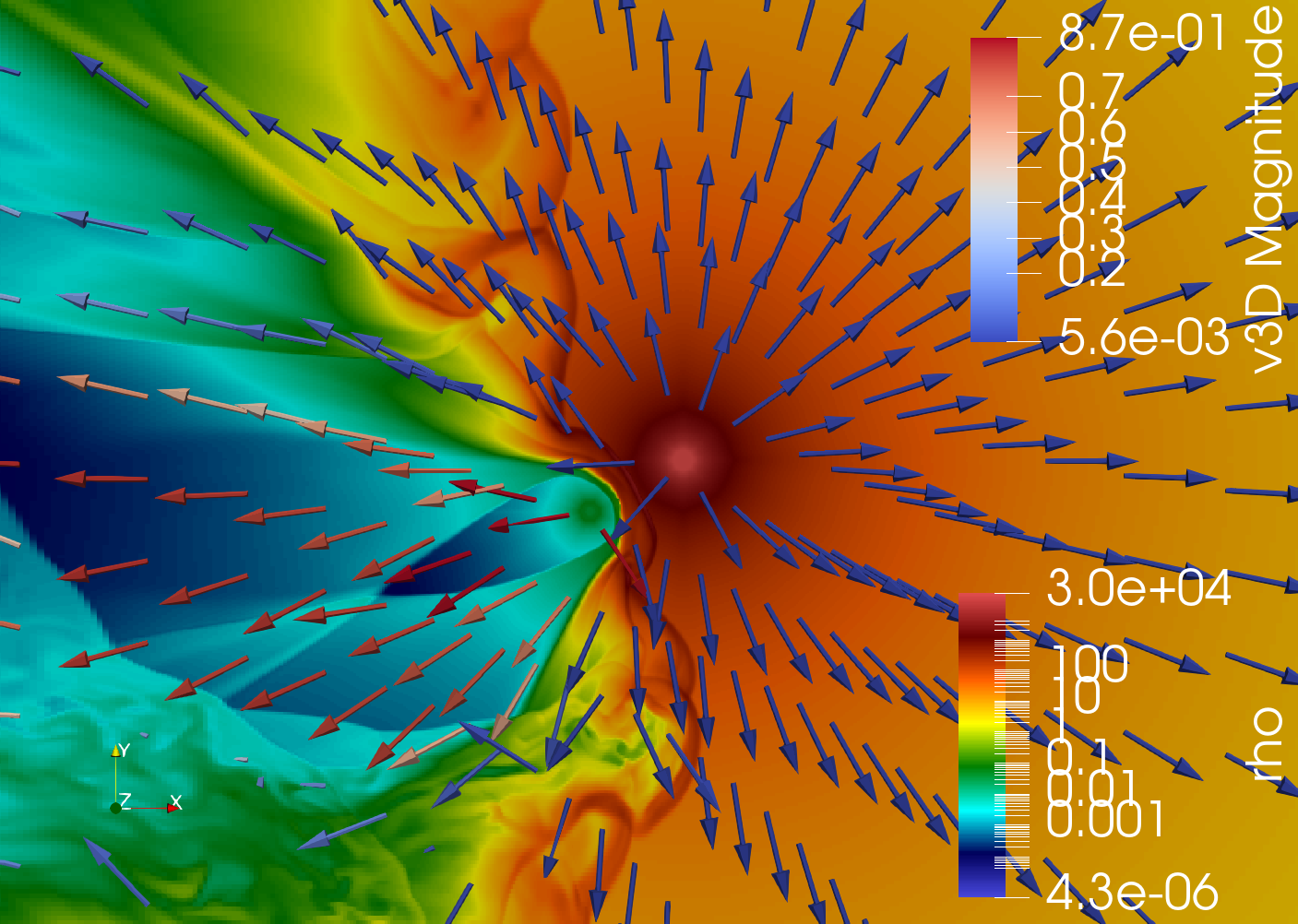}
   \includegraphics[width=88mm]{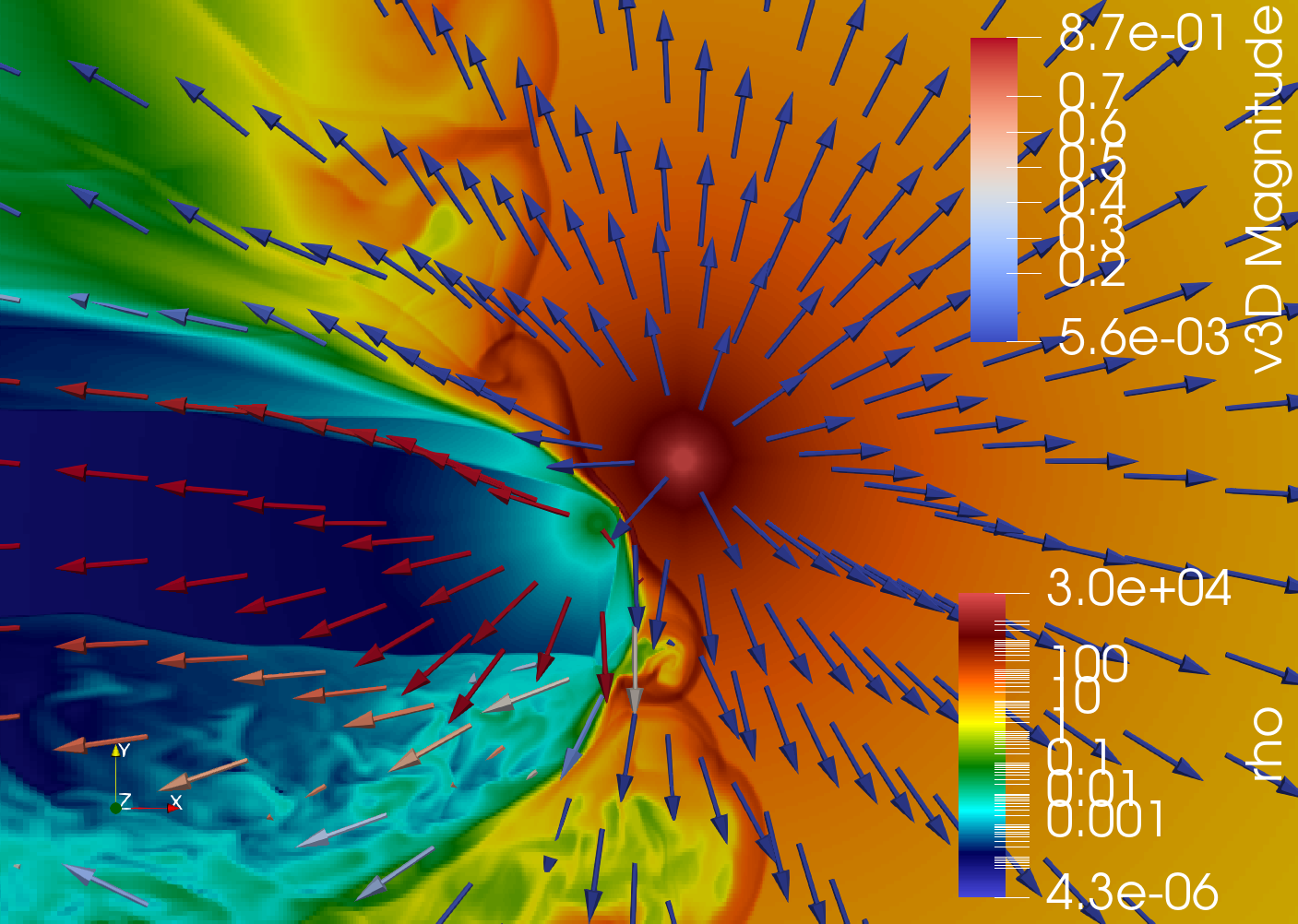}
   \includegraphics[width=88mm]{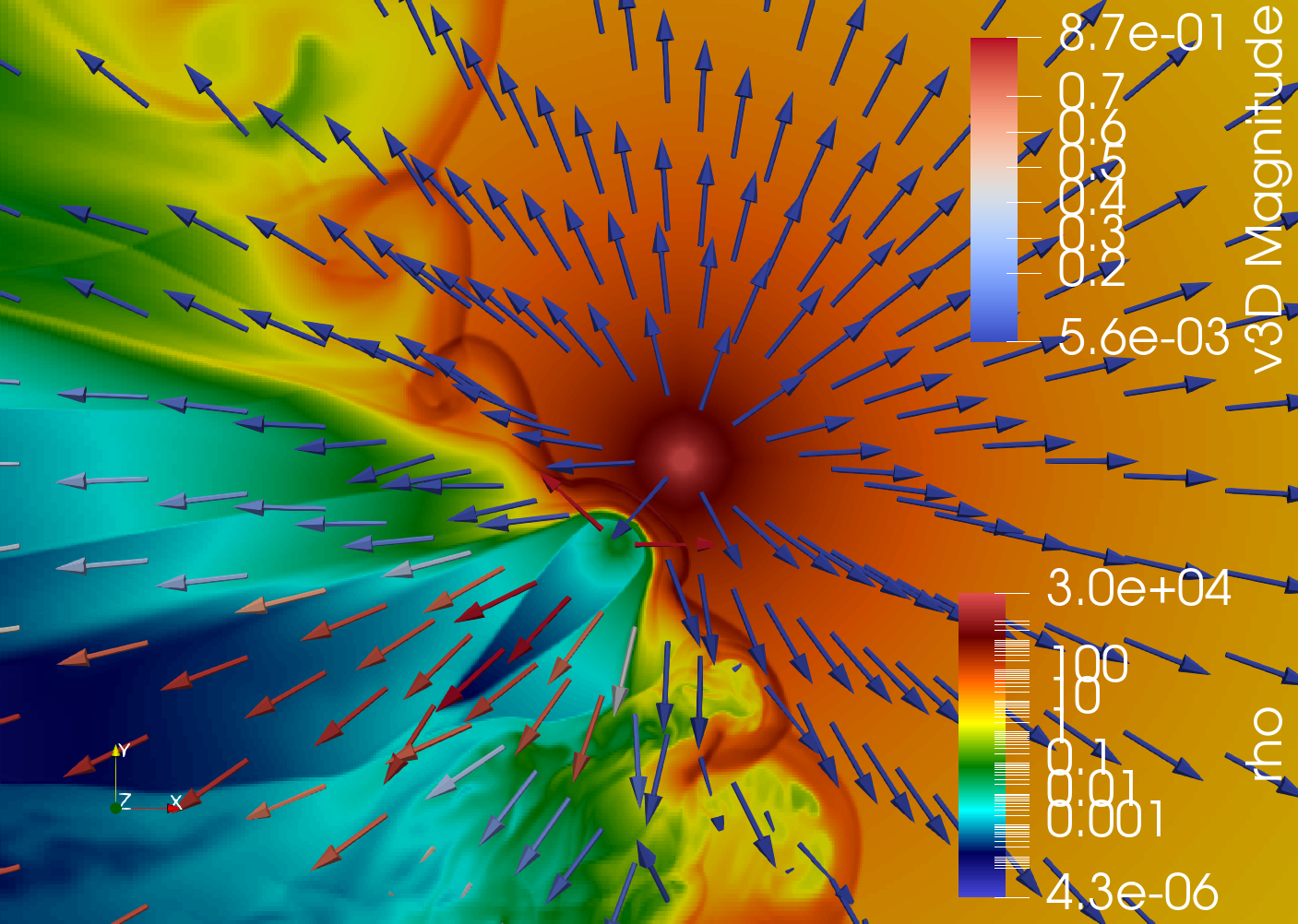}
   \includegraphics[width=88mm]{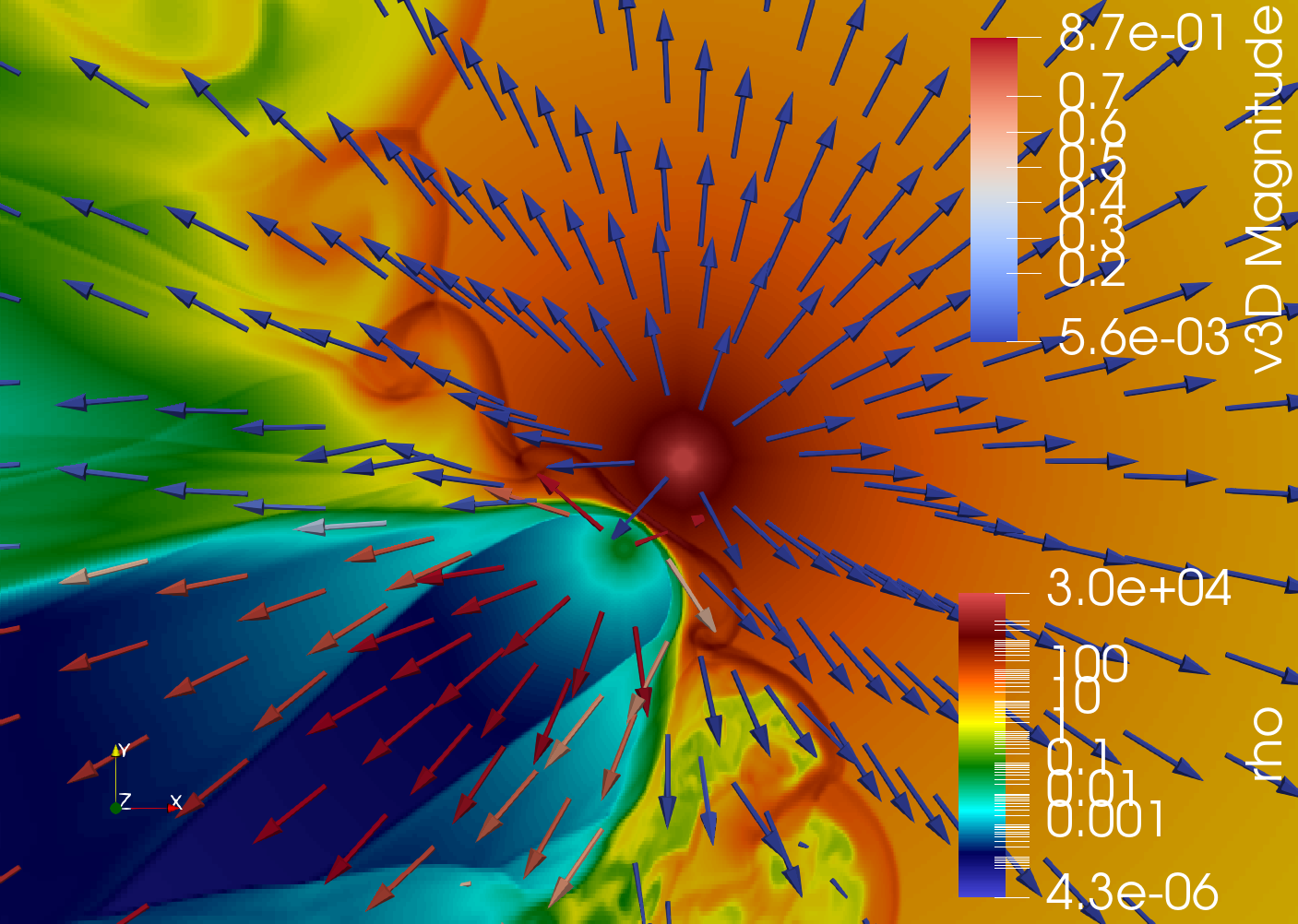}
   \caption{Colored density maps with colored arrows representing the modulus of the 3-velocity vector for the 3D model (3Dn30e10) for cuts on the orbital plane, at orbital phases $\phi = 0.5435 $ (top left), $\phi = 0.575 $ (top right), $\phi = 0.5812 $ (middle left), $\phi = 0.6033 $ (middle right), $\phi = 0.6467 $ (bottom left), and $\phi = 0.6576$ (bottom right).}
    \label{fig:zoomrhov3D}
    \end{figure*}    

\begin{figure}
   \centering 
   \includegraphics[width=\hsize]{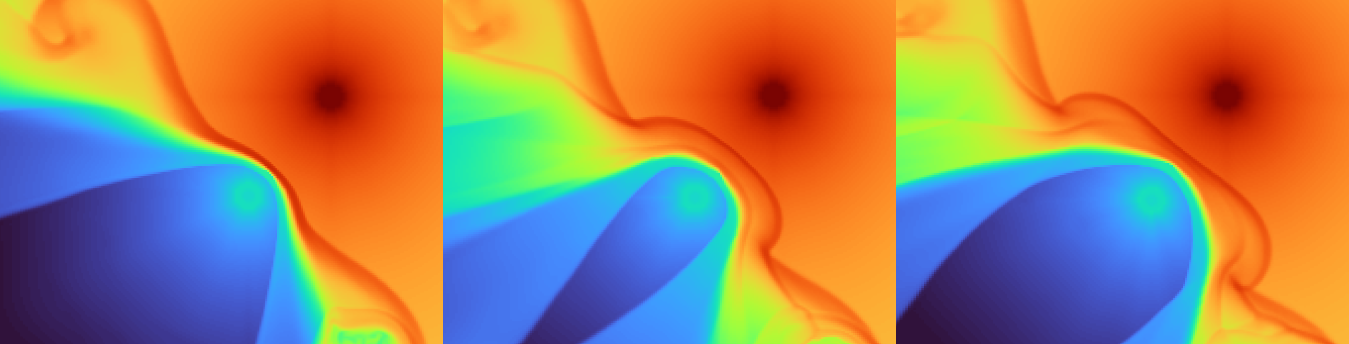}
   \caption{Zoom-in of three snapshots of the cycle followed by the interaction structure: weak pulsar wind (left), enhanced pulsar wind (middle), and the end of energy injection (right).}
   \label{fig:cycle}
\end{figure}

   \begin{figure}
   \centering
   \includegraphics[width=44mm]{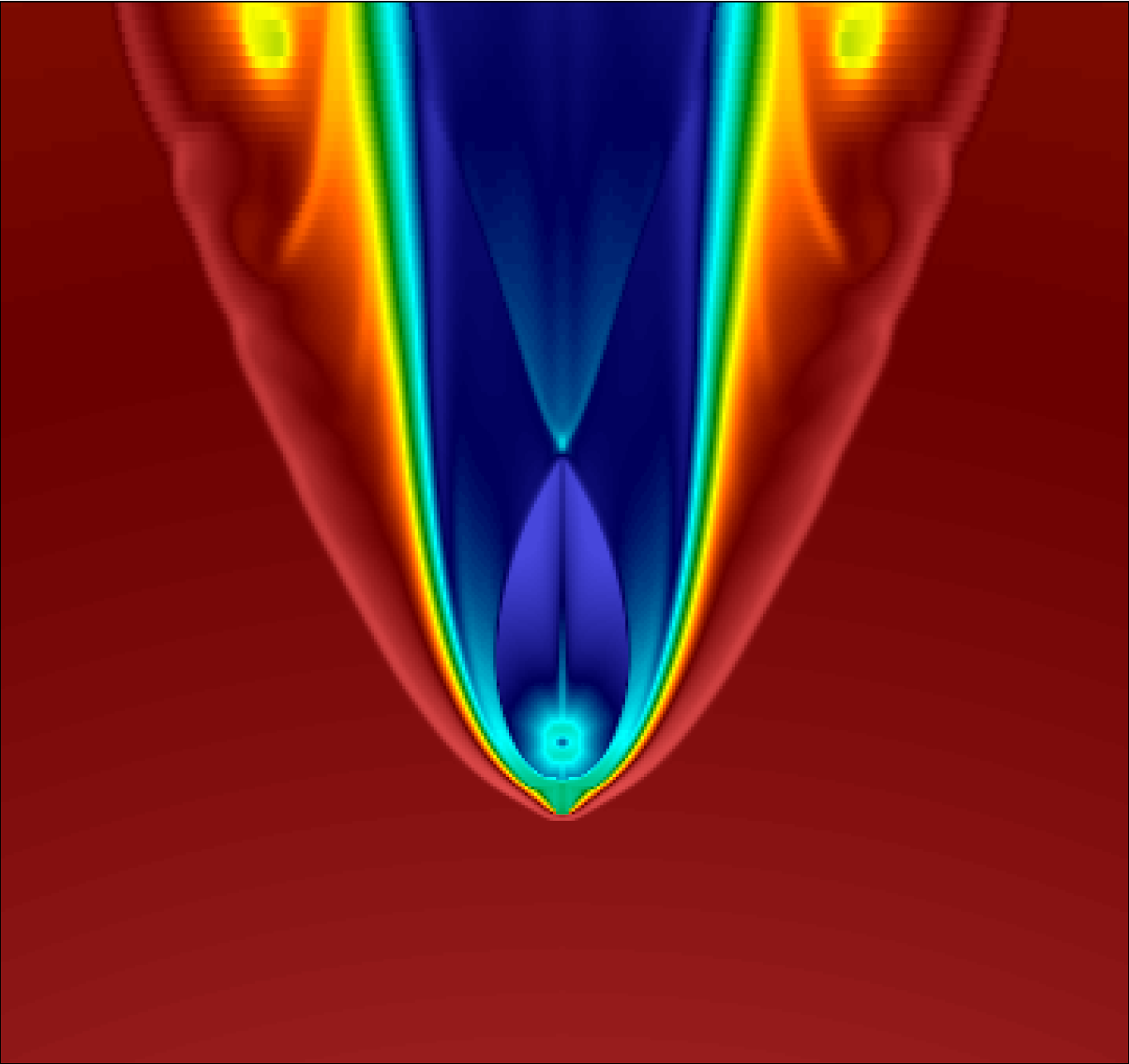}
   \includegraphics[width=44mm]{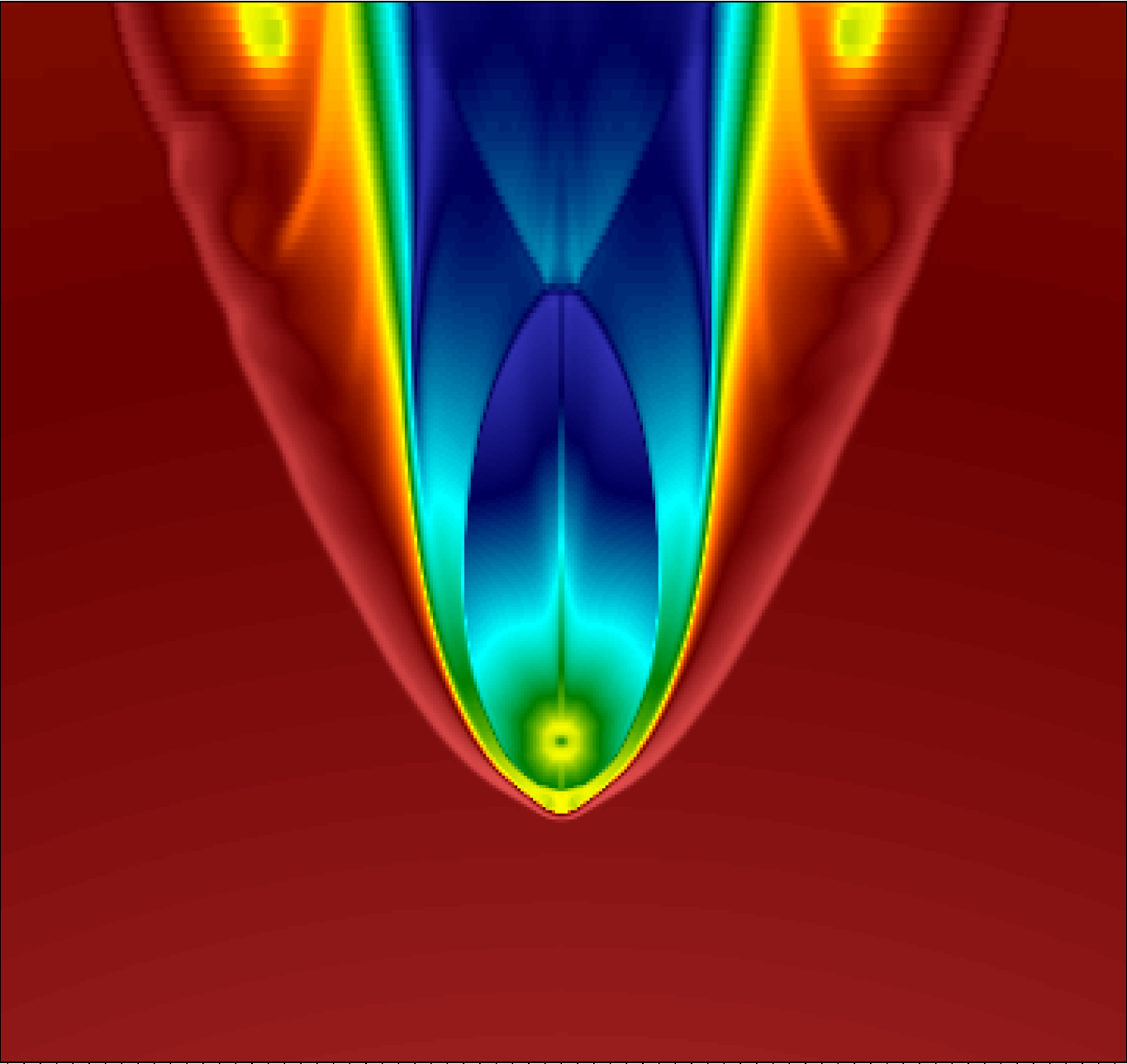}
   \includegraphics[width=44mm]{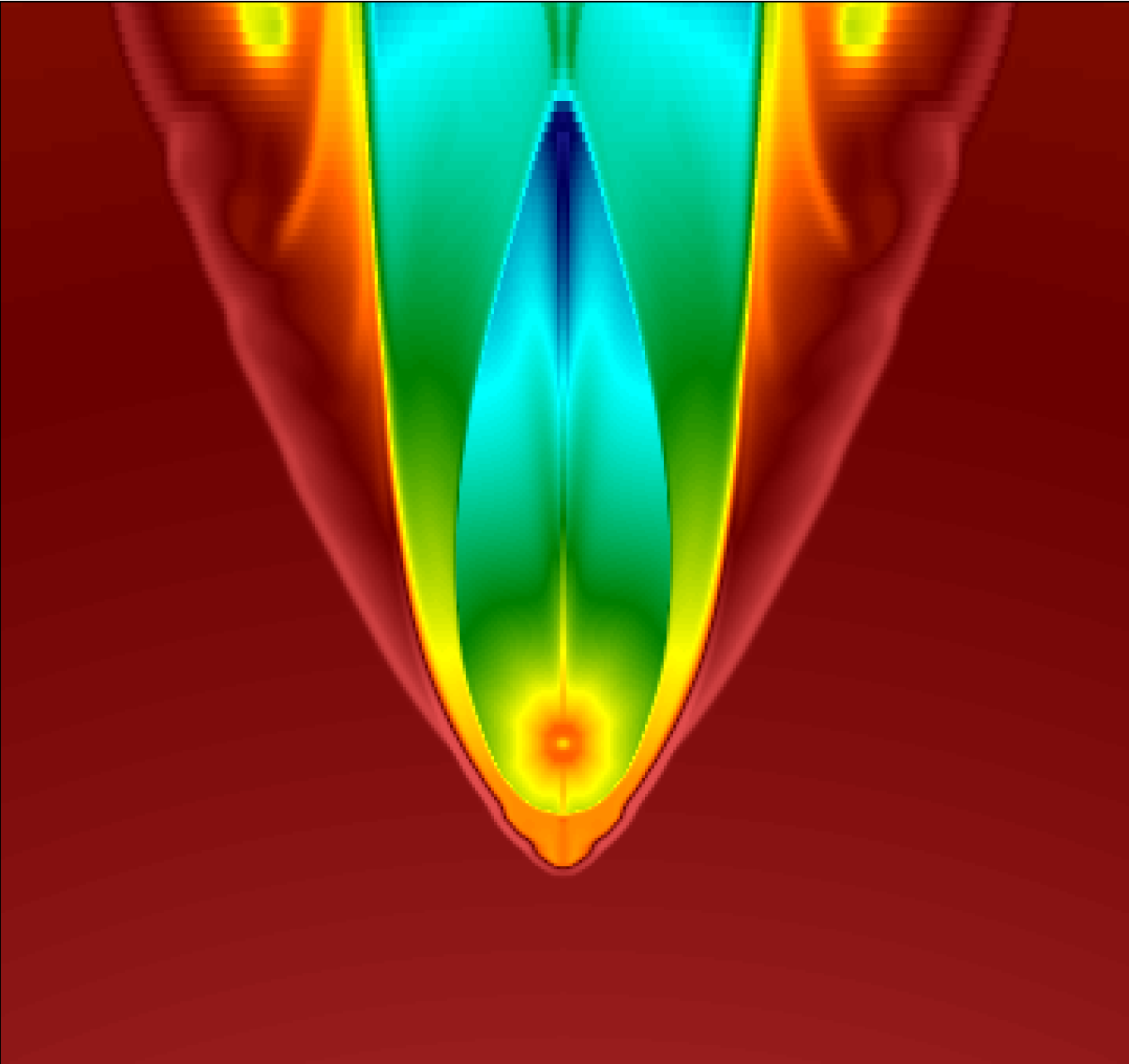}
   \includegraphics[width=44mm]{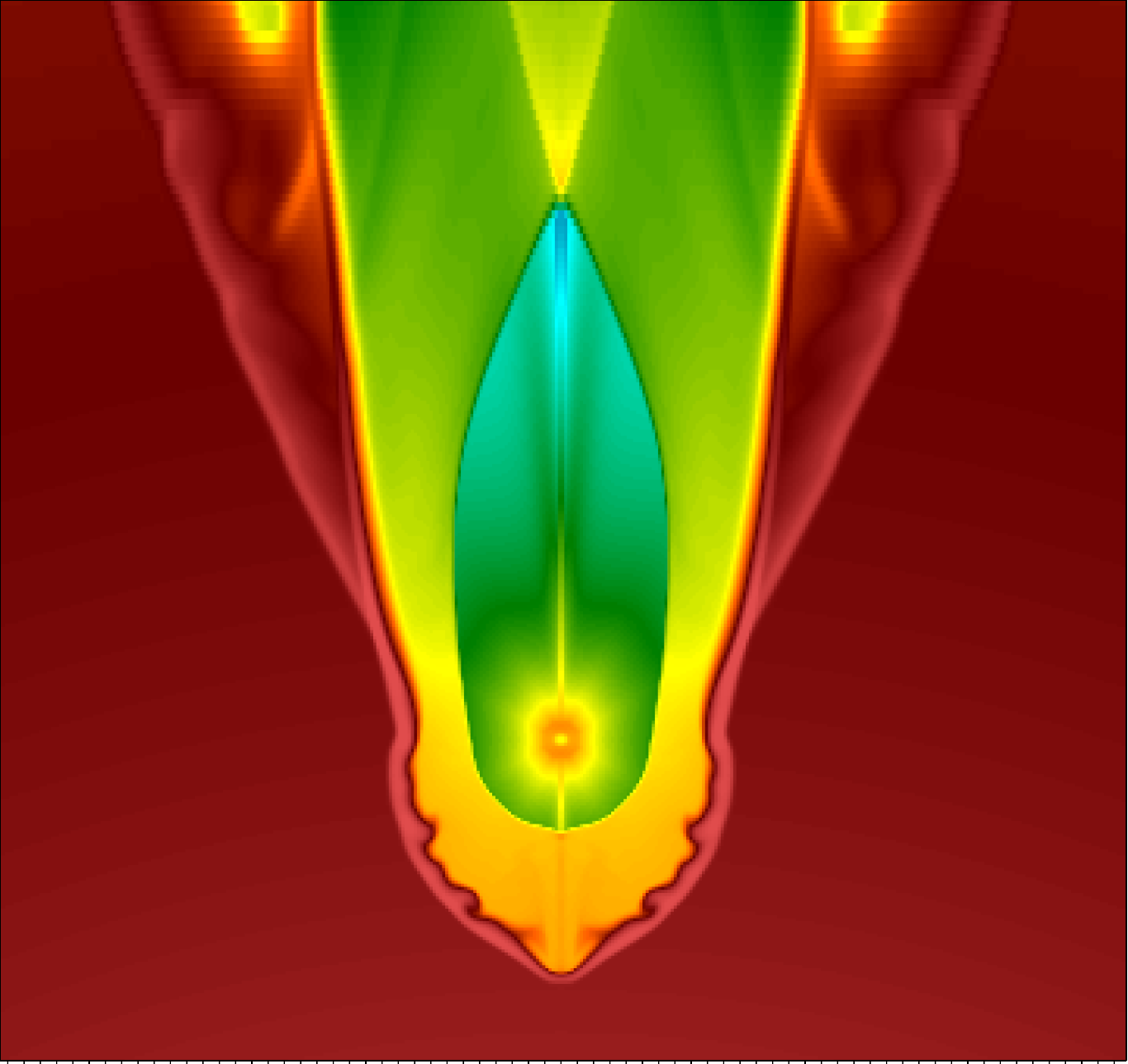}
   \includegraphics[width=44mm]{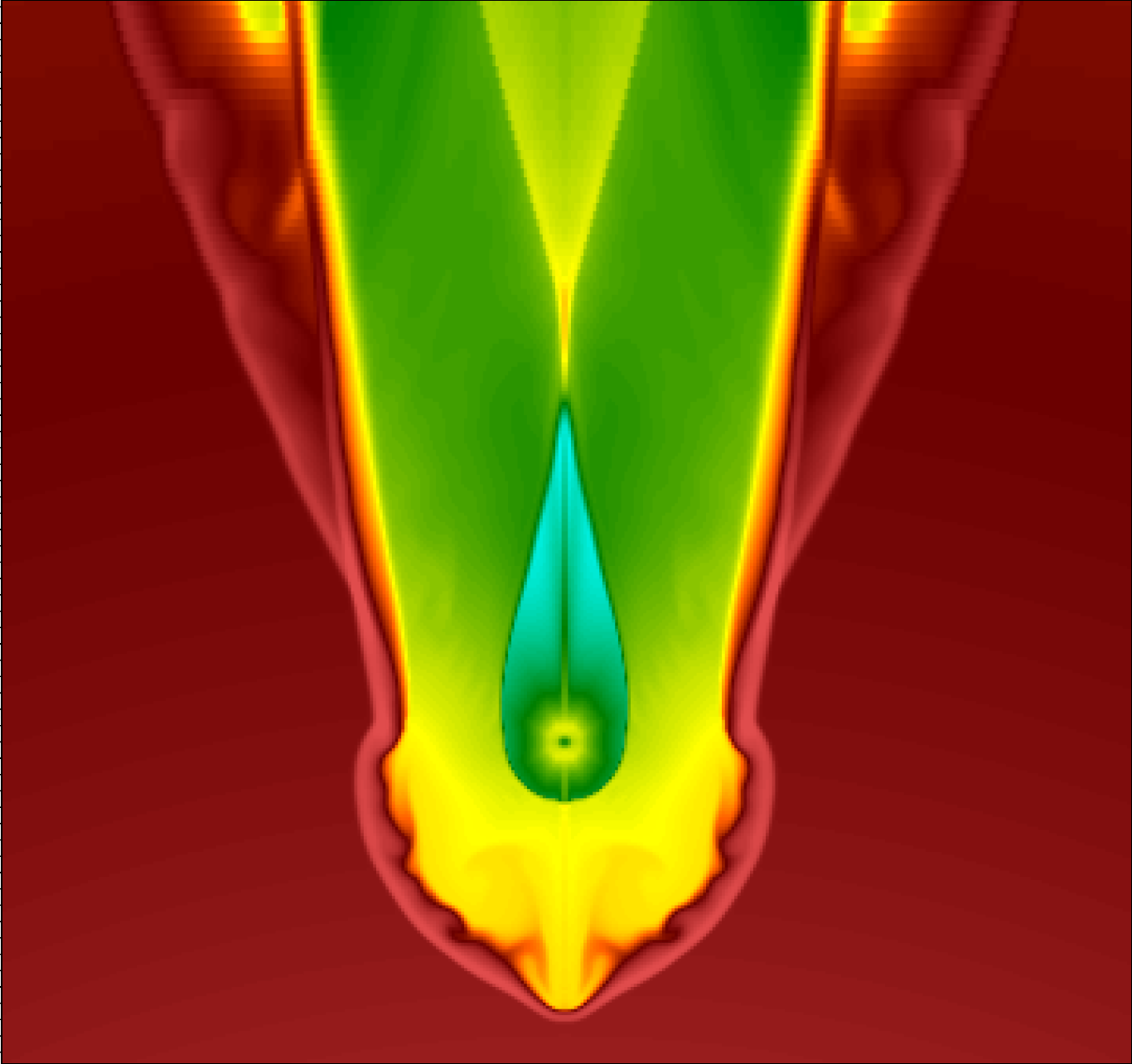}
   \includegraphics[width=44mm]{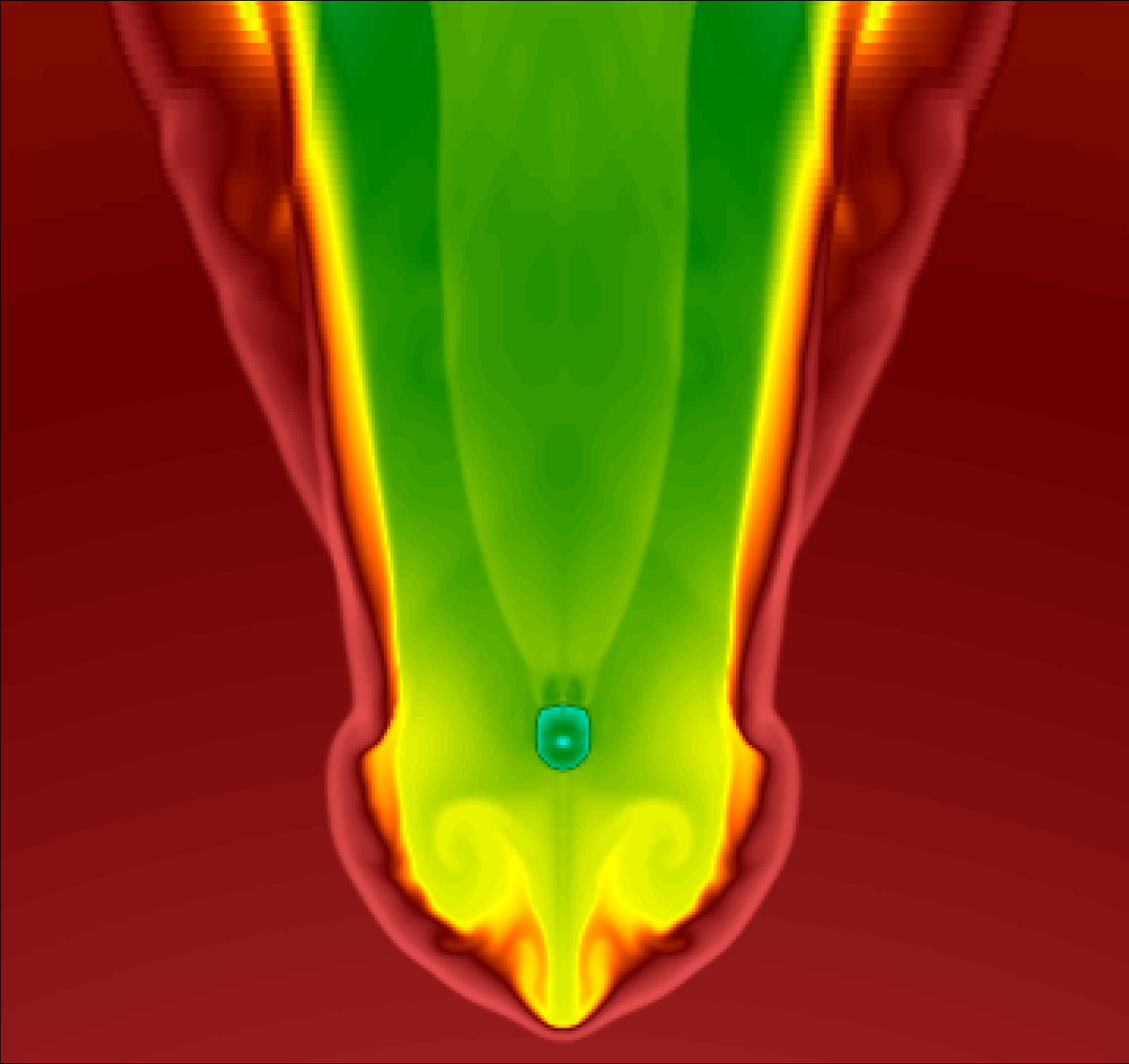}
   \caption{Colored density maps obtained from an axisymmetric simulation showing qualitatively the evolution of a flare, from top left to bottom right: 0 (weak wind), 48 (flare starts), 88 (flare continuation), 141 (end of flare), 166 (pulsar wind approaches magnetosphere), and 186~s (flare about to resume) from the beginning of the flare. The grid size is $\approx 0.65\,a$; the magnetosphere radius would be $\approx 0.02\,a$ in this case.}
    \label{axi}
    \end{figure}  

We carried out simulations for four 2D models and one 3D model in Cartesian coordinates including orbital motion, and one simulation for one 2D axisymmetric model. In the 2D Cartesian models, we varied the number of flaring periods during the orbit ($N_{fl}=10,30,90$) for an orbit-averaged power of the flaring component alone of <$\dot E_{fl}$>$=2.3$<$\dot E_{sd}$>, and for $N_{fl} = 30$, we varied the flaring period power $\times 3$ (i.e., <$\dot E_{fl}$>$=6.8$<$\dot E_{sd}$>). For higher flaring period rates, the interaction structure becomes smoother. In the case 2Dn90e10, with $N_{fl}=90$, the leading edge of the shock is the most uniform (see Fig.~\ref{fig:rhov2D}, left top panel). The result of multiple flaring periods can be seen most clearly in the radial shock fronts that form on the trailing side of the interaction structure. Non-flare solutions are also shown, nfeta0.05 and nfeta0.5, which illustrate two cases with a strong difference in $\eta$, 0.05 (weak wind) and 0.5 (enhanced wind), but without the effect of the flaring periods (left and right bottom panels in Fig.~\ref{fig:rhov2D}). The main differences between the flare and non-flare solutions are: 1) the free pulsar wind zone (FPWZ), which is significantly smaller in the cases with flares, and 2) the spiral tail is significantly more loaded by the stellar wind in those cases. Point 2 is best illustrated by comparing the density of the shocked pulsar wind in the cases with flares and those without. Due to heavy mass-loading, the shocked pulsar wind in the spiral arm is slower in the cases with flares. 

For the cases 2Dn30e10 and 2Dn10e10, the individual flaring periods are more powerful and more separated in time compared to 2Dn90e10 (top right, middle left, and middle right panels in Fig.~\ref{fig:rhov2D}, respectively). As a result, the impacts of individual flaring periods become clearly visible in the density structure of both the leading and the trailing sides of the interaction structure. Increasing the orbit-averaged flaring period power, as seen comparing the 2Dn30e10 and the 2Dn30e30 models (top right and middle right panels in Fig.~\ref{fig:rhov2D}, respectively), leads to a larger two-wind interaction structure, while the FPWZ size decreases when increasing this power. 

In the 3D simulations (model 3Dn30e10; Fig.~\ref{fig:rhov3D}), we switched on the flaring periods for short periods of time in a series of five flaring periods to mimic the flaring period impact within a fraction of the orbit (they occur for one sixth of the orbit), as the higher computational demands in 3D prevented us from probing the effect of the recurrent flaring periods for as long a time as in 2D. As we know from previous work \citep[see, e.g.,][]{2015A&A...577A..89B}, the results in 3D are more turbulent than in 2D in general, so the flares enhance turbulence further. 
The impact of the flaring periods on the overall two-wind interaction structure can be better seen in Fig.~\ref{fig:zoomrhov3D}, although the limitations of the simulations prevented us from reproducing in detail the evolution of the pulsar wind termination shock; this is exemplified by a zoom-in of the interaction structure for the 3D simulations shown in Fig.~\ref{fig:cycle}, where the pulsar wind termination shock does not actually reach the magnetospheric scales. On the other hand, in the 2D axisymmetric simulations presented in Fig.~\ref{axi}, which do not account for orbital motion but probe more realistically the spatial and temporal scales close to the NS, the cycle of weak-enhanced-weak pulsar wind was better captured. The series of events that the simulations show can be described as follows. Initially, the magnetospheric flares are not active, and the weak pulsar wind termination shock can reach close to the NS, mimicking the situation when this shock is about to touch the magnetosphere (Fig.~\ref{axi}, top left\footnote{In reality, the interaction would be unlikely to correspond to two colliding supersonic winds, as it does here, but the simulation was set with such a configuration to simplify the numerical setup.}). When this happens, the flaring period resumes (Fig.~\ref{axi}, top right). Due to its high density, the shocked stellar wind reacts slowly to the power-enhanced pulsar wind and the shocked pulsar wind stays in contact with the magnetosphere, still exciting the magnetospheric activity for some time (Fig.~\ref{axi}, middle left). Eventually, as the shocked stellar wind is pushed outward, the pulsar wind termination shock moves away from the magnetosphere, and the extra energy injection stops (Fig.~\ref{axi}, middle right). At that point, the weakening of the pulsar wind allows the shocked pulsar wind to quickly propagate inward, toward the magnetosphere (Fig.~\ref{axi}, bottom panels), the system again becoming ready to restart the cycle. Using 2D axisymmetric simulations on small scales more systematically to better inform simulations on large scales (which here deal in a simplified manner with those small scales) is to be tackled in future studies. 

\section{Summary and discussion} \label{disc} 

In the previous section, we
described how magnetospheric energy can be recurrently injected in the two-wind
interaction region, and we illustrated this process with RHD simulations. Some
caveats were raised in Sect.~\ref{intro}, as the pulsar wind enhanced by magnetar
flares is expected to be magnetized and not spherically symmetric
\citep[e.g.,][]{2023MNRAS.524.6024S}. However, these differences from a spherical
RHD wind are alleviated by the fact that other common factors (stellar wind,
the nonsteady nature of the shocked flows, orbital motion, etc.) also play a strong
role in the evolution of the interaction structure. Moreover, strong relativistic
waves are expected between the magnetosphere and the stellar wind during magnetar
flares, which can lead to a more spherical shocked pulsar wind, and to a reduction
in the wind magnetization. 
Assessing all this requires detailed future analysis,
and better connecting processes taking place on different spatial and temporal scales. Putting these caveats aside, in what follows some aspects of the
proposed scenario are discussed; namely, the regime in which the stellar wind
interacts with the NS surroundings, the nonthermal energy origin and the age of
the source, the latter being relevant as no recent SNR has been found near
LS~5039. The article finishes with a few remarks regarding the nonthermal emitter. In all the expressions below, we adopted the convention $A_{\xi}=A/10^\xi$, with $A$ being in cgs units but for $\dot{M}$, which is in units of solar masses per year.

\subsection{NS-medium interaction regime}

A NS interacting with a stellar wind can have two regimes in LS~5039: (1) ejector and (2) georotator \citep[or propeller; see below;][]{1975wsfx.book.....I,1992ans..book.....L}, which are discussed below. The state of the system is characterized by several radii: The distance at which pulsar and stellar winds would be in ram pressure balance\footnote{In a steady two-wind interaction (ejector) regime, this distance is roughly similar but somewhat larger than the pulsar wind termination shock radius.}:
\begin{equation}
R_s=a\frac{\eta^{1/2}}{1+\eta^{1/2}}\approx 3.5\times 10^{10}a_{12.3}\eta_{-3.5}^{1/2}\,{\rm cm}\,\,\,({\rm assuming}\,\eta\ll 1)\,;
\label{eq:rs}
\end{equation}
the light cylinder radius, 
\begin{equation}
R_{LC}=cP/2\pi\approx 3.8\times 10^{10}P_{0.9}\,{\rm cm}; 
\label{eq:rlc}
\end{equation}
the Alfvenic radius at the balance location between the stellar wind ram pressure and the dipolar magnetic pressure of the NS magnetosphere:
\begin{equation}
 R_A= \left(\frac{a^2\mu^2}{2\dot{M}v_w}\right)^{1/6}\approx
2.7\times 10^{10}\frac{a^{1/3}_{12.3}\mu^{1/3}_{33}}{\dot{M}^{1/6}_{-6.4}v^{1/6}_{w,8.3}}\,{\rm cm},\label{eq:ra}
\end{equation}
where $\mu=B_{dp}R_{NS}^3$ is the dipolar NS magnetic moment, and $B_{dp}$ the dipolar magnetic field at the NS surface; and the gravitational capture radius:
\begin{equation}
    R_g = \frac{2G M_{NS}}{v_w^2}\approx 9.3\times10^9\,v_{w,8.3 }^{-2} \mbox{ cm},
    \label{eq:racc}
\end{equation}
for $M_{NS}=1.4$~M$_\odot$ and a wind speed normalization close to, but a bit lower than, its velocity at infinity of $\approx 2.4\times 10^8$~cm~s$^{-1}$ \citep{cas05}, as we took into account the compactness of the system. For $B_{dp}\gtrsim 4\times 10^{13}$~G (and taking $R_{NS}=10^6$~cm), one has $R_g<R_A$, which is the reason we focus on the georotator regime hereafter. Nevertheless, the propeller regime could still be compatible with the triggering of magnetic flares. We did not account for a multipolar magnetic field (see below) when deriving $R_A$ as such a field drops much faster than $B_{dp}$. 

The induced magnetospheric flares could be possible either in a marginal ejector regime, or in the georotator regime. Let us discuss first the ejector regime.

\subsubsection{Ejector regime}

The ejector regime takes place if $R_s>R_{LC}$. From the dipole formula, the spin-down power becomes
\begin{equation}
    \dot{E}_{sd} = \frac{\mu^2\Omega^4}{c^3}(1+\sin(\alpha)^2)\sim 1.5\times 10^{34}\frac{\mu_{33}^2}{P_{0.9}^{4}}\,{\rm erg~s}^{-1},
    \label{eq:Edot}
\end{equation}
where $\Omega=2\pi/P$. For simplicity, in Eq.~(\ref{eq:Edot}) we neglected the dependence on the angle between between the magnetic momentum and the rotation axis, which is $\sim(1+\sin(\alpha)^2)$ (see, for numerical force-free simulations, e.g., \citealt{1999A&A...349.1017B,2006ApJ...648L..51S}, and for the analytical expression $(2/3)\sin(\alpha)^2$, e.g., the review by \citealt{2010PhyU...53.1199B}), and we simplified it to $\sim 1$, but it does not change our conclusions significantly. The marginal ejector regime would correspond then to $R_s$ temporarily reaching $R_{LC}$. Taking $R_s=R_{LC}$, one can derive a minimum dipolar magnetic field for the ejector regime: 
\begin{equation}
    B_{cr}\approx\frac{c^3}{8\pi^3}\frac{P^3\sqrt{\dot{M}v_w}}{R_{NS}^3a} \approx 
    2\times 10^{15} \frac{P_{0.9}^3\dot{M}_{-6.4}^{1/2} v_{w,8.3}^{1/2}}{R_{NS,6}^{3} a_{12.3}}  \quad \mbox{G},
    \label{eq:Bcr}
\end{equation}
where we assumed $\eta\ll 1$. In the marginal ejector regime, the magnetospheric flares would occur when $B_{dp}\sim B_{cr}$. If $B_{dp}=B_{cr}$, the dipole spin-down time would be
\begin{equation}
    T_{sd}\approx\frac{1}{24}\frac{IP^3c^3}{\pi^2\mu^2}\approx\frac{8 \pi^4}{3c^3}\frac{a^2 I}{P^3\dot{M}v_w} \approx
    500  \frac{I_{45}a^2_{12.3}}{P_{0.9}^3\dot{M}_{-6.4}v_{w,8.3}}\quad \mbox{yr,}
    \label{eq:tsd}
\end{equation}
where $I\approx10^{45}$g~cm$^2$ is the NS moment of inertia. 

The $T_{sd}$ value obtained assuming $B_{dp}=B_{cr}$ is similar to the $\tau\sim 500$~yr inferred from the X-ray data by \cite{yon20} and \cite{mak23} for the pulsar spin-down. Under the same assumption, the spin-down power, $\dot{E}_{sd}$, is $\sim 7\times10^{34}$~erg~s$^{-1}$, which is also close to the value inferred by \cite{yon20}, and significantly below $L_{NT}\sim 10^{36}$~erg~s$^{-1}$ in LS~5039. This spin-down power would correspond to the weak pulsar wind phase in the simulations, that is, between flaring periods. 
We recall that in the Cartesian simulations including orbital motion, $\eta=0.05$ was adopted due to resolution limitations, making the simulation results qualitative, and in reality $\eta$ would be much smaller, as in the axisymmetric simulation. For $\dot{E}_{sd}\sim 7\times 10^{34}$~erg~s$^{-1}$, 
$\dot{M}\sim 4\times10^{-7}$~M$_\odot$~yr$^{-1}$, and $v_w\sim 2\times 10^8$~cm~s$^{-1}$, one obtains $\eta\sim 5\times10^{-4}$.

\subsubsection{Georotator regime}

For $B_{dp}<B_{cr}$, which implies that $R_s<R_{LC}$, the system is in the georotator regime and the stellar wind directly interacts with the NS magnetosphere, and the maximum spin-down power associated with this interaction can be derived as follows. First, the mass rate of the stellar wind intercepted by the NS magnetosphere is obtained:
\begin{equation}
\dot{M}_{NS}\approx\dot{M}\frac{R_{A}^2}{4a^2}=\left(\frac{\mu^2\dot{M}^2}{2^7v_wa^4}\right)^{1/3},
    \label{eq:mdns}
\end{equation}
which, following \citep{2020arXiv200411474L}, allows for the maximum power dissipated on the magnetosphere boundary to be derived, as the incoming flow is forced to corotate with the NS magnetic field\footnote{We note that this estimate is not very robust and needs verification by multidimensional magnetohydrodynamical modeling.}:
\begin{equation}
\dot{E}_{Gr}\approx\frac{\pi^2}{2^{2/3}P^2}\left(\frac{\mu^4\dot{M}}{v_w^2a^2}\right)^{1/3}\approx 2.5\times 10^{34}\,\frac{\mu_{32}^{4/3}\dot{M}^{1/3}_{-6.4}}{P_{0.9}^{2}v_{w,8.3}^{2/3}a_{12.3}^{2/3}}\,{\rm erg~s}^{-1}.
    \label{eq:lpr}
\end{equation}
This value is similar to the $\dot{E}_{sd}$ value obtained above, so the small amount of energy released in this regime when no flares happen may also be considered equivalent to the weak pulsar wind phase in the simulations.

From the maximum power estimate given in Eq.~(\ref{eq:lpr}), in the georotator regime the spin-down time is
\begin{equation}
    T_{sd,Gr} \gtrsim 2^{5/3}\frac{I v_w^{2/3}a^{2/3}}{\mu^{4/3}\dot{M}^{1/3}}\approx 400\,\frac{I_{45}v_{w,8.3}^{2/3}a_{12.3}^{2/3}}{\mu_{32}^{4/3}\dot{M}_{-6.4}^{1/3}}\,{\rm yr}.
    \label{eq:taupr}
\end{equation}
Taking $T_{sd,Gr}=\tau\sim 500$~yr from \cite{yon20}, one can estimate the minimum dipolar NS magnetic moment as ($\tau$ units are seconds: 500~yr $\approx 10^{10.2}$~s)
\begin{equation}
    \mu_{min}\approx 8\times10^{31} 
    \frac{I_{45}^{3/4} v_{w,8.3}^{1/2} a_{12.3}^{1/2}}{\tau_{10.2}^{3/4}\dot{M}_{-6.4}^{1/4}}
    \quad \mbox{G cm}^3,
    \label{eq:mumin}
\end{equation}
so the minimum dipolar NS magnetic field is
\begin{equation}
    B_{dp,min} \approx 8\times10^{13} 
    \frac{I_{45}^{3/4} v_{w,8.3}^{1/2} a_{12.3}^{1/2}}{R_{NS,6}^3\tau_{10.2}^{3/4}\dot{M}_{-6.4}^{1/4}}
    \quad \mbox{G}.
    \label{eq:bmin}
\end{equation}
Adopting $B_{dp}=B_{cr}$, $R_g$ is $\sim 0.3\,R_{A}$ for the source specific parameters, whereas taking $B_{dp}=B_{dp,min}$, $R_g$ becomes $\sim 0.8 R_{A}$, still marginally in the georotator regime. Importantly, the lower value of $B_{dp}$ in the latter case can lead to a much longer ejector phase; for $\mu=\mu_{min}$,
\begin{equation}
    T_{sd}(\mu_{min})\approx 4\times10^3\frac{P_{0.3}^3 \tau_{10.2}^{3/2} \dot{M}_{-6.4}^{1/2}}{v_{w,8.3}a_{12.3}I_{45}^{1/2}}\, \mbox{yr},
    \label{eq:tauEJmax}
\end{equation}
where $P$ has been normalized now to 2~s to account for pulsar rotation
braking while still fulfilling $R_s>R_{LC}$ (ejector). If one sets $B_{cr}=B_{dp,min}$, one can obtain the longest pulsar period for which this condition is still fulfilled:
\begin{equation}
    P_{max}\approx 2.8\frac{I_{45}^{1/4}a_{12.3}^{1/2}}{\tau_{10.2}^{1/4} \dot{M}_{-6.4}^{1/4}}\, \mbox{ s}
    \label{eq:Ppmax}
.\end{equation}
Substituting it into Eq.~(\ref{eq:tauEJmax}) yields the duration of the ejector phase corresponding to $P_{max}$:
\begin{equation}
    T_{sd,max}\approx 1.1\times10^4 
    \frac{
    \tau_{10.2}^{3/4}a_{12.3}^{1/2}I_{45}^{1/4}}{v_{w,8.3}\dot{M}_{-6.4}^{1/4}}\,\mbox{ yr}.
    \label{eq:tauSDmax}
\end{equation}
An important conclusion is reached here: if the source were in the georotator regime between flares, and not marginally in the ejector regime as first discussed, the previous ejector phase could have lasted for $\sim 10^4$~yr due to the much lower $B_{dp}$ and an initial pulsar period of $\sim P_{max}$.

\subsection{Nonthermal energy origin}

The magnetic energy of the NS can be estimated as
\begin{equation}
    E_{mag}\approx\chi R_{NS}^3B_{NS}^2\approx 1.7\times10^{47} R_{NS,6}^3 B_{NS, 15}^2\mbox{ erg},
    \label{eq:emag}
\end{equation}
with the constant $\chi$ depending on the field topology ($1/6$ for a dipole field, as adopted by us, and slightly smaller for a higher multipole field).
Here, $B_{NS}$ includes magnetic field components beyond the dipolar one, as magnetar-like activity requires of reconnection events that occur when the magnetic field has high multipoles. This multipolar NS magnetic field can be characterized through an amplification factor, $x_B$, with respect to $B_{dp}$, which can be significantly larger than 1: $B_{NS}=x_B B_{dp}\gg B_{dp}$. The multipolar magnetic field can be generated through the dynamo mechanism; for instance, when the compact object formed \citep[see, e.g.,][and references there in]{Landau1980Classical, 1993ApJ...408..194T, 2015Natur.528..376M,2020SciA....6.2732R, 2023MNRAS.524.6024S,2025A&A...695A.183B,2025NatAs...9..541I}.

The dissipation of the NS magnetic field takes place on a timescale of
\begin{equation}
    \tau_{mag} \lesssim \frac{E_{mag}}{L_{NT}} \sim  5\times 10^3\frac{R_{NS,6}^3 B_{NS, 15}^2}{L_{NT,36}} \mbox{ yr},
    \label{eq:taummax}
\end{equation}
where we took into account that not all the energy released in the magnetic flares goes into nonthermal emission, as some of this energy goes to make work on the environment or heat the NS, or to nonthermal particles that do not radiate efficiently. In the case of $B_{NS}\sim B_{dp,min}$ and the georotator regime, the magnetic dissipation timescale would be $\lesssim 30$~yr, which is completely implausible. To sustain $L_{NT}$ for $\sim 10^4$~yr for instance, at least $B_{NS}\sim 1.4\times 10^{15}$~G is needed, meaning that $x_B\gtrsim 20$. 

\begin{figure}
\includegraphics[width=\hsize]{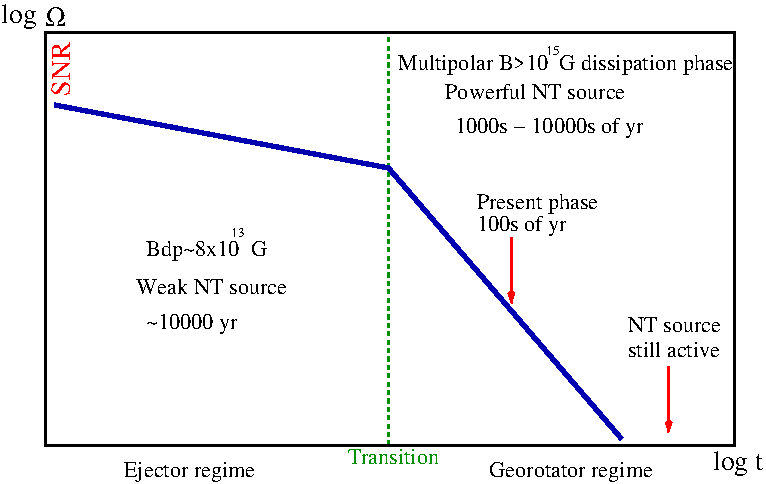}
\caption{Sketch of the georotator scenario evolution, showing the pulsar angular velocity evolution with time with the ejector and the georotator (present) phases and a dashed green line marking the transition.}
\label{sketch}
\end{figure}

A marginal ejector case with $B_{NS}\sim B_{dp}\sim B_{cr}$, and thus $x_B\sim 1$, could sustain $L_{NT}$ for a few centuries until the georotator regime were reached, and the subsequent georotator regime, albeit powerful, would evolve very fast (in $\sim 10$~yr, as $B_{dp}$ should still be $\sim B_{cr}$ to feed $L_{NT}$, see Eq.~\ref{eq:taupr}), all of this implying a very young SNR. Therefore, a present georotator regime with a large $\chi_B$ (i.e., with a relatively small $B_{dp}$) would be favored for LS~5039. In addition, in this case, even if the present georotator phase is just a few centuries old, and the NS slows down its rotation significantly during that timescale, this regime may last much longer with substantial nonthermal activity, say for $\sim 10^4$~yr. Thus, such a type of source can be an active accelerator for a relatively long time. A sketch of the scenario based on a present georotator regime for LS~5039 is shown in Fig.~\ref{sketch}.

\subsection{Lack of SNR evidence}

As has been shown, in the georotator regime, the total time elapsed since the NS was formed can be $\gg\tau\sim 500$~yr, the age from \cite{yon20}. There is first the period in which the pulsar was still in the ejector regime, which as explained may have lasted for $\sim 10^4$~yr depending on the initial period and magnetic field. Then, in the georotator regime, the system may have evolved for some time until reaching the present stage, with a current decaying time of several centuries. Nevertheless, regardless of the georotator decaying time, since the multipolar magnetic field can be strong even when the NS rotation slows down, the actual magnetic field dissipation phase can last much longer (e.g., $\tau_{mag}\sim 10^4$~yr). From all this, the SNR may have occurred long ago and become by now already inconspicuous and difficult to detect. 

Another, less likely, possibility for the absence of observational evidence of a SNR associated with LS~5039 could be a very small ejecta mass (e.g., up to $0.1 M_\odot$) in the case of a white dwarf (WD) turning into a NS \citep[see eg.][]{1992ApJ...391..228W,2025MNRAS.541.1649K}. Such a scenario is unlikely as the rate of such a kind of SN explosion, particularly in high-mass binaries, is very low. Such a scenario may have occurred if the WD progenitor had transferred most of its mass to the secondary star, nowadays an O6.5~V star \citep{cas05}, preventing its initial direct collapse into a NS. In this quite improbable scenario, the NS might be quite young but no SNR would be visible.

\subsection{The nonthermal emitter}

The proposed scenario is not very different from the standard colliding wind scenario \citep[e.g.,][]{tav97}, as the former still features a relativistic outflow leaving the magnetosphere, but its power source is expected to be more irregular. Unfortunately, the simulations presented do not allow us properly connect the processes close to the NS with those on larger scales, and further work is required to properly characterize the development of an enhanced pulsar wind, and how it propagates and terminates, all of this in the context of the evolution of the whole interaction structure. It is known that the acceleration of electrons up to the required energies in LS~5039 is difficult \citep{kha08,bos25}, and only an adequate characterization of that region can assess its ability to produce the particles with the required energies. Further downstream, the differences between the two scenarios are qualitatively smaller, although the more perturbed nature of the shocked flows due to magnetospheric flaring may suggest that particle acceleration on the scales of the spiral formation is stronger in that case. We emphasize that highly relativistic flows can still form outside the magnetosphere when strong flares occur, which can be a basic ingredient \citep[as in the model proposed by][]{der12,bos25} for the production of ultra-high-energy photons such as those observed in LS~5039 \citep{alf25}. The overall nonthermal emitter is, however, likely to spread over a much larger region \citep[e.g.,][]{kha08,2012A&A...544A..59B}. We finish by noting that the timescales inherent to the scenario explored here are diverse, from the light-crossing timescale of the magnetosphere (seconds) and the stellar wind crossing time of the interaction region (minutes to hours), to the orbital period (days).

\begin{acknowledgements}
The authors want to thank the referee for a useful report that helped to improve the clarity of the manuscript.
VB-R acknowledges financial support from the Spanish Ministry of Science and Innovation under grant
PID2022-136828NB-C41/AEI/10.13039/501100011033/ERDF/EU and through the Mar\'ia de Maeztu award to the ICCUB (CEX2024-001451-M), and from the Generalitat de Catalunya through grant 2021SGR00679. VB-R is Correspondent Researcher of CONICET, Argentina, at the IAR. Part of this work was supported by the Russian Science Foundation grant 23-22-00385 (sections 2 and 4) and BASIS foundation grant \#24-1-2-25-1 (sections 1 and 3). The calculations were performed on CFCA XC50 cluster of the National Astronomical Observatory of Japan (NAOJ) and RIKEN HOKUSAI Bigwaterfall.   
\end{acknowledgements}

\bibliographystyle{aa}
\bibliography{lsmag} 

\begin{thebibliography}{41}
\expandafter\ifx\csname natexlab\endcsname\relax\def\natexlab#1{#1}\fi

\bibitem[{{Alfaro} {et~al.}(2025){Alfaro}, {Araya}, {Arteaga-Vel{\'a}zquez},
  {Rojas}, {Ayala Solares}, {Babu}, {Bangale}, {Belmont-Moreno}, {Bernal},
  {Caballero-Mora}, {Capistr{\'a}n}, {Carrami{\~n}ana}, {Casanova}, {Cotti},
  {Cotzomi}, {Couti{\~n}o de Le{\'o}n}, {Depaoli}, {Desiati}, {Di Lalla}, {Diaz
  Hernandez}, {Dingus}, {DuVernois}, {Engel}, {Ergin}, {Espinoza}, {Fan},
  {Fang}, {Garc{\'\i}a-Gonz{\'a}lez}, {Goksu}, {Gonzalez Mu{\~n}oz},
  {Gonz{\'a}lez}, {Gonz{\'a}lez}, {Goodman}, {Groetsch}, {Harding},
  {Hern{\'a}ndez-Cadena}, {Herzog}, {Hinton}, {Huang}, {Hueyotl-Zahuantitla},
  {H{\"u}ntemeyer}, {Kaufmann}, {Kieda}, {Lara}, {Lee}, {Le{\'o}n Vargas},
  {Linnemann}, {Longinotti}, {Luis-Raya}, {Malone}, {Martinez},
  {Mart{\'\i}nez-Castro}, {Matthews}, {Miranda-Romagnoli}, {Morales-Soto},
  {Moreno}, {Mostaf{\'a}}, {Najafi}, {Nellen}, {Nisa}, {Omodei}, {Ponce},
  {P{\'e}rez Araujo}, {P{\'e}rez-P{\'e}rez}, {Rho}, {Rodriguez Parra},
  {Rosa-Gonz{\'a}lez}, {Roth}, {Salazar}, {Salazar-Gallegos}, {Sandoval},
  {Schneider}, {Schwefer}, {Serna-Franco}, {Smith}, {Son}, {Springer},
  {Tibolla}, {Tollefson}, {Torres}, {Torres-Escobedo}, {Turner}, {Varela},
  {Wang}, {Wang}, {Watson}, {Wu}, {Yu}, {Yun-C{\'a}rcamo}, {Zhou}, {de
  Le{\'o}n}, \& {HAWC Collaboration}}]{alf25}
{Alfaro}, R., {Araya}, M., {Arteaga-Vel{\'a}zquez}, J.~C., {et~al.} 2025,
  \apjl, 987, L42

\bibitem[{{Barkov} {et~al.}(2024){Barkov}, {Kalinin}, \&
  {Lyutikov}}]{2024PASA...41...48B}
{Barkov}, M., {Kalinin}, E., \& {Lyutikov}, M. 2024, \pasa, 41, e048

\bibitem[{{Barkov} \& {Popov}(2022)}]{2022MNRAS.515.4217B}
{Barkov}, M.~V. \& {Popov}, S.~B. 2022, \mnras, 515, 4217

\bibitem[{{Barkov} {et~al.}(2022){Barkov}, {Sharma}, {Gourgouliatos}, \&
  {Lyutikov}}]{bar22}
{Barkov}, M.~V., {Sharma}, P., {Gourgouliatos}, K.~N., \& {Lyutikov}, M. 2022,
  \apj, 934, 140

\bibitem[{{Barr{\`e}re} {et~al.}(2025){Barr{\`e}re}, {Guilet}, {Raynaud}, \&
  {Reboul-Salze}}]{2025A&A...695A.183B}
{Barr{\`e}re}, P., {Guilet}, J., {Raynaud}, R., \& {Reboul-Salze}, A. 2025,
  \aap, 695, A183

\bibitem[{{Beloborodov}(2020)}]{2020ApJ...896..142B}
{Beloborodov}, A.~M. 2020, \apj, 896, 142

\bibitem[{{Beskin}(2010)}]{2010PhyU...53.1199B}
{Beskin}, V.~S. 2010, Physics Uspekhi, 53, 1199

\bibitem[{{Bogovalov}(1999)}]{1999A&A...349.1017B}
{Bogovalov}, S.~V. 1999, \aap, 349, 1017

\bibitem[{{Bogovalov} {et~al.}(2019){Bogovalov}, {Khangulyan}, {Koldoba},
  {Ustyugova}, \& {Aharonian}}]{2019MNRAS.490.3601B}
{Bogovalov}, S.~V., {Khangulyan}, D., {Koldoba}, A., {Ustyugova}, G.~V., \&
  {Aharonian}, F. 2019, \mnras, 490, 3601

\bibitem[{{Bosch-Ramon} {et~al.}(2012){Bosch-Ramon}, {Barkov}, {Khangulyan}, \&
  {Perucho}}]{2012A&A...544A..59B}
{Bosch-Ramon}, V., {Barkov}, M.~V., {Khangulyan}, D., \& {Perucho}, M. 2012,
  \aap, 544, A59

\bibitem[{{Bosch-Ramon} {et~al.}(2015){Bosch-Ramon}, {Barkov}, \&
  {Perucho}}]{2015A&A...577A..89B}
{Bosch-Ramon}, V., {Barkov}, M.~V., \& {Perucho}, M. 2015, \aap, 577, A89

\bibitem[{{Bosch-Ramon} \& {Khangulyan}(2025)}]{bos25}
{Bosch-Ramon}, V. \& {Khangulyan}, D. 2025, \aap, 700, A162

\bibitem[{{Bozzo} {et~al.}(2008){Bozzo}, {Falanga}, \& {Stella}}]{boz08}
{Bozzo}, E., {Falanga}, M., \& {Stella}, L. 2008, \apj, 683, 1031

\bibitem[{{Casares} {et~al.}(2005){Casares}, {Rib{\'o}}, {Ribas}, {Paredes},
  {Mart{\'\i}}, \& {Herrero}}]{cas05}
{Casares}, J., {Rib{\'o}}, M., {Ribas}, I., {et~al.} 2005, \mnras, 364, 899

\bibitem[{{Collmar} \& {Zhang}(2014)}]{col14}
{Collmar}, W. \& {Zhang}, S. 2014, \aap, 565, A38

\bibitem[{{Derishev} \& {Aharonian}(2012)}]{der12}
{Derishev}, E.~V. \& {Aharonian}, F.~A. 2012, in American Institute of Physics
  Conference Series, Vol. 1505, High Energy Gamma-Ray Astronomy: 5th
  International Meeting on High Energy Gamma-Ray Astronomy, ed. F.~A.
  {Aharonian}, W.~{Hofmann}, \& F.~M. {Rieger} (AIP), 402--405

\bibitem[{{Igoshev} {et~al.}(2025){Igoshev}, {Barr{\`e}re}, {Raynaud},
  {Guilet}, {Wood}, \& {Hollerbach}}]{2025NatAs...9..541I}
{Igoshev}, A., {Barr{\`e}re}, P., {Raynaud}, R., {et~al.} 2025, Nature
  Astronomy, 9, 541

\bibitem[{{Illarionov} \& {Syunyayev}(1975)}]{1975wsfx.book.....I}
{Illarionov}, A.~F. \& {Syunyayev}, R.~A. 1975, {Why are there so few X-ray
  stars?}

\bibitem[{{Kargaltsev} {et~al.}(2023){Kargaltsev}, {Hare}, {Volkov}, \&
  {Lange}}]{kar23}
{Kargaltsev}, O., {Hare}, J., {Volkov}, I., \& {Lange}, A. 2023, \apj, 958, 79

\bibitem[{{Khangulyan} {et~al.}(2008){Khangulyan}, {Aharonian}, \&
  {Bosch-Ramon}}]{kha08}
{Khangulyan}, D., {Aharonian}, F., \& {Bosch-Ramon}, V. 2008, \mnras, 383, 467

\bibitem[{{Khangulyan} {et~al.}(2022){Khangulyan}, {Barkov}, \&
  {Popov}}]{2022ApJ...927....2K}
{Khangulyan}, D., {Barkov}, M.~V., \& {Popov}, S.~B. 2022, \apj, 927, 2

\bibitem[{{Kuroda} {et~al.}(2025){Kuroda}, {Kawaguchi}, \&
  {Shibata}}]{2025MNRAS.541.1649K}
{Kuroda}, T., {Kawaguchi}, K., \& {Shibata}, M. 2025, \mnras, 541, 1649

\bibitem[{Landau \& Lifshitz(1980)}]{Landau1980Classical}
Landau, L.~D. \& Lifshitz, E.~M. 1980, The Classical Theory of Fields, 4th edn.
  (Butterworth-Heinemann)

\bibitem[{{Li}(2005)}]{2005JCoPh.203..344L}
{Li}, S. 2005, Journal of Computational Physics, 203, 344

\bibitem[{{Lipunov}(1992)}]{1992ans..book.....L}
{Lipunov}, V.~M. 1992, {Astrophysics of Neutron Stars}

\bibitem[{{Lorimer} {et~al.}(2007){Lorimer}, {Bailes}, {McLaughlin},
  {Narkevic}, \& {Crawford}}]{2007Sci...318..777L}
{Lorimer}, D.~R., {Bailes}, M., {McLaughlin}, M.~A., {Narkevic}, D.~J., \&
  {Crawford}, F. 2007, Science, 318, 777

\bibitem[{{Lyutikov} {et~al.}(2020{\natexlab{a}}){Lyutikov}, {Barkov}, {Route},
  {Balsara}, {Garnavich}, \& {Littlefield}}]{2020arXiv200411474L}
{Lyutikov}, M., {Barkov}, M., {Route}, M., {et~al.} 2020{\natexlab{a}}, arXiv
  e-prints, arXiv:2004.11474

\bibitem[{{Lyutikov} {et~al.}(2020{\natexlab{b}}){Lyutikov}, {Barkov}, \&
  {Giannios}}]{2020ApJ...893L..39L}
{Lyutikov}, M., {Barkov}, M.~V., \& {Giannios}, D. 2020{\natexlab{b}}, \apjl,
  893, L39

\bibitem[{{Makishima} {et~al.}(2023){Makishima}, {Uchida}, {Yoneda}, {Enoto},
  \& {Takahashi}}]{mak23}
{Makishima}, K., {Uchida}, N., {Yoneda}, H., {Enoto}, T., \& {Takahashi}, T.
  2023, \apj, 959, 79

\bibitem[{{Mignone} {et~al.}(2007){Mignone}, {Bodo}, {Massaglia}, {Matsakos},
  {Tesileanu}, {Zanni}, \& {Ferrari}}]{2007ApJS..170..228M}
{Mignone}, A., {Bodo}, G., {Massaglia}, S., {et~al.} 2007, \apjs, 170, 228

\bibitem[{{Mold{\'o}n} {et~al.}(2012){Mold{\'o}n}, {Rib{\'o}}, {Paredes},
  {Brisken}, {Dhawan}, {Kramer}, {Lyne}, \& {Stappers}}]{mol12}
{Mold{\'o}n}, J., {Rib{\'o}}, M., {Paredes}, J.~M., {et~al.} 2012, \aap, 543,
  A26

\bibitem[{{M{\"o}sta} {et~al.}(2015){M{\"o}sta}, {Ott}, {Radice}, {Roberts},
  {Schnetter}, \& {Haas}}]{2015Natur.528..376M}
{M{\"o}sta}, P., {Ott}, C.~D., {Radice}, D., {et~al.} 2015, \nat, 528, 376

\bibitem[{{Porth} {et~al.}(2014){Porth}, {Komissarov}, \&
  {Keppens}}]{2014MNRAS.438..278P}
{Porth}, O., {Komissarov}, S.~S., \& {Keppens}, R. 2014, \mnras, 438, 278

\bibitem[{{Raynaud} {et~al.}(2020){Raynaud}, {Guilet}, {Janka}, \&
  {Gastine}}]{2020SciA....6.2732R}
{Raynaud}, R., {Guilet}, J., {Janka}, H.-T., \& {Gastine}, T. 2020, Science
  Advances, 6, 2732

\bibitem[{{Sharma} {et~al.}(2023){Sharma}, {Barkov}, \&
  {Lyutikov}}]{2023MNRAS.524.6024S}
{Sharma}, P., {Barkov}, M.~V., \& {Lyutikov}, M. 2023, \mnras, 524, 6024

\bibitem[{{Spitkovsky}(2006)}]{2006ApJ...648L..51S}
{Spitkovsky}, A. 2006, \apjl, 648, L51

\bibitem[{{Tavani} \& {Arons}(1997)}]{tav97}
{Tavani}, M. \& {Arons}, J. 1997, \apj, 477, 439

\bibitem[{{Thompson} \& {Duncan}(1993)}]{1993ApJ...408..194T}
{Thompson}, C. \& {Duncan}, R.~C. 1993, \apj, 408, 194

\bibitem[{{Weng} {et~al.}(2022){Weng}, {Qian}, {Wang}, {Torres}, {Papitto},
  {Jiang}, {Xu}, {Li}, {Yan}, {Liu}, {Ge}, \& {Yuan}}]{wen22}
{Weng}, S.-S., {Qian}, L., {Wang}, B.-J., {et~al.} 2022, Nature Astronomy, 6,
  698

\bibitem[{{Woosley} \& {Baron}(1992)}]{1992ApJ...391..228W}
{Woosley}, S.~E. \& {Baron}, E. 1992, \apj, 391, 228

\bibitem[{{Yoneda} {et~al.}(2020){Yoneda}, {Makishima}, {Enoto}, {Khangulyan},
  {Matsumoto}, \& {Takahashi}}]{yon20}
{Yoneda}, H., {Makishima}, K., {Enoto}, T., {et~al.} 2020, \prl, 125, 111103

\end{thebibliography}

\end{document}